\def\doublecolumn{1} 
\if 1\doublecolumn
  \documentclass[10pt,journal]{IEEEtran}
\else
  \documentclass[12pt, journal, onecolumn]{IEEEtran}
\fi

\def\blind{1} 

\usepackage[latin9]{inputenc}
\usepackage{url}
\usepackage{array}
\usepackage{makecell} 
\usepackage{float}
\usepackage{dsfont}
\usepackage{mathtools}
\usepackage{amsmath}
\usepackage{amsthm}
\usepackage{amssymb}

\usepackage{graphicx}
\usepackage{setspace}
\usepackage{color}

\usepackage{bm}
\usepackage{cases}
\usepackage{bbm}

\usepackage{hhline}

\usepackage{epsfig}
\usepackage{cite}

\usepackage{array}
\usepackage{enumerate}
\usepackage{enumitem}

\usepackage{times}

\usepackage[framemethod=tikz]{mdframed}
\definecolor{cccolor}{rgb}{1,1,1}

\usepackage{multicol,multirow,adjustbox}
\usepackage{booktabs}
\usepackage[caption=false,font=footnotesize,labelformat=simple]{subfig}

\newtheorem{theorem}{Theorem}
\newtheorem{lemma}{Lemma}
\newtheorem{remark}{Remark}

\DeclareMathOperator*{\argmax}{argmax}

\newcommand{\eg}[0]{\textit{e.g.}}
\newcommand{\ie}[0]{\textit{i.e.}}

\usepackage{xr-hyper} 
\usepackage[bookmarks=false]{hyperref}

\usepackage{soul}

\definecolor{hlcolor}{HTML}{ED7117}

\soulregister\cite7
\soulregister\ref7
\soulregister\cref7
\soulregister\eqref7
\soulregister\pageref7
\soulregister\url{7}

\usepackage[capitalize]{cleveref}
\crefname{section}{Sec.}{Secs.}
\Crefname{section}{Section}{Sections}
\crefname{table}{Tab.}{Tabs.}
\Crefname{table}{Table}{Tables}
\Crefname{assumption}{Assumption}{Assumptions}

\makeatletter
\def\@IEEEsectpunct{.\ \,}
\def\paragraph{\@startsection{paragraph}{4}{\z@}{1.2ex plus 1.1ex minus 0.5ex}%
{0ex}{\normalfont\normalsize\bfseries}}
\makeatother

\usepackage{pifont,xspace}
\mdfdefinestyle{customspacing}{
  outerlinewidth=1pt,
  linewidth=1pt,
  middlelinewidth=0pt,
  backgroundcolor=yellow!60,
  linecolor=orange,
  innerrightmargin=4pt,
  innerleftmargin=4pt,
  innertopmargin=4pt,
  innerbottommargin=4pt
}

\newenvironment{tablehl*}[1][htbp]
  {\begin{table*}[#1]\begingroup\mdfsetup{skipabove=-8pt,skipbelow=-8pt}
  \begin{mdframed}[style=customspacing]}
  {\end{mdframed}\endgroup\end{table*}}

\mdfdefinestyle{figurecustomspacing}{
  outerlinewidth=1pt,
  linewidth=1pt,
  middlelinewidth=0pt,
  backgroundcolor=yellow!60,
  linecolor=orange,
  innerrightmargin=2pt,  
  innerleftmargin=2pt,   
  innertopmargin=2pt,    
  innerbottommargin=2pt  
}

\newenvironment{figurehl*}[1][htbp]
  {\begin{figure*}
  [#1]\begingroup\mdfsetup{skipabove=-8pt,skipbelow=-12pt}
  \begin{mdframed}[style=figurecustomspacing]}
  {\end{mdframed}\endgroup\end{figure*}}

\usepackage[linesnumbered,ruled,vlined]{algorithm2e}
\makeatletter
\newcommand{\removelatexerror}{\let\@latex@error\@gobble}
\makeatother

\DontPrintSemicolon

\SetKwComment{Comment}{\color{green!50!black}// }{}

\SetKwProg{Function}{function}{}{}

\SetKwProg{Init}{Initialize}{}{}
\SetKwProg{Inp}{Input}{}{}
\SetKwProg{Out}{Output}{}{}

\SetKwComment{Comment}{\color{green!50!black}// }{}

\newcommand{\var}{\texttt}

\SetKwProg{Function}{function}{}{}

\newcounter{problem}          
\newcounter{save@equation}    
\newcounter{save@problem}     

\makeatletter
\newenvironment{problem}
 {
  \setcounter{problem}{\value{save@problem}}%
  \setcounter{save@equation}{\value{equation}}%
  \let\c@equation\c@problem   
  \subequations               
 }
 {
  \endsubequations            
  \setcounter{save@problem}{\value{equation}}%
  \setcounter{equation}{\value{save@equation}}%
 }
\makeatother

\def\argmax{\mathop{\mathrm{argmax}}}

\def\b0{{\pmb{0}}} 

\DeclareFixedFont{\ttb}{T1}{txtt}{bx}{n}{12} 
\DeclareFixedFont{\ttm}{T1}{txtt}{m}{n}{12}  

\usepackage{color}
\definecolor{deepblue}{rgb}{0,0,0.5}
\definecolor{deepred}{rgb}{0.6,0,0}
\definecolor{deepgreen}{rgb}{0,0.5,0}

\usepackage{listings}

\newcommand\pythonstyle{\lstset{
language=Python,
basicstyle=\ttm\footnotesize,
morekeywords={self},              
keywordstyle=\ttb\color{deepblue}\footnotesize,
emph={MyClass,__init__},          
emphstyle=\ttb\color{deepred}\footnotesize,    
stringstyle=\color{deepgreen}\footnotesize,
frame=tblr,                         
showstringspaces=false,
breaklines=true
}}

\lstnewenvironment{python}[1][]
{
\pythonstyle
\lstset{#1}
}
{}

\newcommand\pythoninline[1]{{\pythonstyle\lstinline!#1!}}

\begin{document}

\title{
Joint Optimization of User Association and Resource Allocation for Load Balancing With Multi-Level Fairness
}

\if 1\blind
\author{
Jonggyu~Jang,~\IEEEmembership{Member,~IEEE}, 
Hyeonsu~Lyu,~\IEEEmembership{Student Member,~IEEE},
David~J.~Love,~\IEEEmembership{Fellow,~IEEE},
and Hyun~Jong~Yang,~\IEEEmembership{Member,~IEEE}
     \thanks{
    J. Jang and D. J. Love are with the Elmore Family School of Electrical and Computer Engineering, Purdue University, West Lafayette, IN 47907 USA (e-mail: \{jang255, djlove\}@purdue.edu).
    H. Lyu is with the Department of Electrical Engineering, Pohang University of Science and Technology (POSTECH), Pohang 37673, Republic of Korea (email: hslyu4@postech.ac.kr).
    H. J. Yang is with the Department of Electrical and Computer Engineering, Seoul National University, Seoul 08826, Republic of Korea (e-mail: hjyang@snu.ac.kr).
    The corresponding authors are David J. Love and Hyun Jong Yang.
    }
}
\else
\author{Anonymous Submission}
\fi

\IEEEtitleabstractindextext{%
\begin{abstract}
    User association, the problem of assigning each user device to a suitable base station, is increasingly crucial as wireless networks become denser and serve more users with diverse service demands.  
    The joint optimization of user association and resource allocation (UARA) is a fundamental issue for future wireless networks, as it plays a pivotal role in enhancing overall network performance, user fairness, and resource efficiency.
    Given the latency-sensitive nature of emerging network applications, network management favors algorithms that are simple and computationally efficient rather than complex centralized approaches. 
    Thus, distributed pricing-based strategies have gained prominence in the UARA literature, demonstrating practicality and effectiveness across various objective functions, \eg, sum-rate, proportional fairness, max-min fairness, and alpha-fairness. 
    While the alpha-fairness frameworks allow for flexible adjustments between efficiency and fairness via a single parameter $\alpha$, existing works predominantly assume a homogeneous fairness context, assigning an identical $\alpha$ value to all users.  
    Real-world networks, however, frequently require differentiated user prioritization due to varying application requirements and latency.
    To bridge this gap, we propose a novel heterogeneous alpha-fairness (HAF) objective function, assigning distinct $\alpha$ values to different users, thereby providing enhanced control over the balance between throughput, fairness, and latency across the network.
    We present a distributed, pricing-based optimization approach utilizing an auxiliary variable framework and provide analytical proof of its convergence to an $\epsilon$-optimal solution, where the optimality gap $\epsilon$ decreases with the number of iterations. 
    Our numerical results demonstrate the effectiveness of the proposed HAF method, highlighting its superior performance relative to conventional homogeneous fairness schemes across multiple performance criteria.
\end{abstract}
}

\maketitle

\section{Introduction}

The emergence of 6G wireless networks is refining the architecture and operational demands of modern communication systems. 
These future networks are expected to accommodate massive user connectivity, ultra-low latency, and intelligent, context-aware resource management~\cite{brinton2025key,andrews20246}. 
In this landscape, the joint optimization of user association and resource allocation (UARA) has become a fundamental challenge. 
User association (UA)---determining the most suitable base station (BS) for each user---is particularly critical in dense, heterogeneous environments where users exhibit diverse demands~\cite{xu2021survey,liu2016user}. 
Addressing these demands often requires resolving inherent trade-offs between \textit{fairness} and \textit{efficiency}: allocating more resources to underperforming users improves fairness but may reduce overall system throughput, whereas prioritizing high-rate users enhances efficiency at the expense of user fairness.
A well-designed UARA strategy must therefore support differentiated resource control, adapting flexibly to the heterogeneous requirements of modern wireless services.


\subsection{Backgrounds}

\paragraph*{Fairness-aware network optimization}

Fairness has long been a central objective in network resource allocation (RA), primarily because of the inefficiencies and user dissatisfaction caused by purely throughput-maximizing strategies. 
Two foundational criteria have shaped the fairness-efficiency trade-off.
The first criterion is \textit{proportional fairness} (PF), introduced by~\cite{kelly1998rate}, which maximizes the sum of logarithmic utilities across users. 
PF is widely recognized for balancing system throughput with fairness. 
The second criterion is \textit{max-min fairness}, which aims to maximize the minimum utility across users~\cite{1095081}, thereby ensuring strong fairness guarantees, especially in resource-constrained or service-critical environments.

To unify these objectives under a generalized mathematical representation, the authors of ~\cite{879343} introduced the concept of \textit{alpha-fairness} ($\alpha$-fairness) where the choice of $\alpha$ incorporates between sum-rate maximization ($\alpha=0$), proportional fairness ($\alpha=1$), and max-min fairness ($\alpha=\infty$).
This generalization has enabled broad adoption of $\alpha$-fairness as a versatile tool for designing multi-objective utility functions.
Theoretically, extensions of $\alpha$-fairness have been explored in wireless environments with axiomatic analysis~\cite{5461911}, multi-objective learning~\cite{ban2024fair}, and distributed resource allocation with incomplete information~\cite{altman2008generalized,altman2009alpha}.

\paragraph*{Distributed pricing-based optimization}

As network scale and complexity increased, centralized solutions became less viable due to signaling overhead and computational demands. 
To tackle the complexity of centralized optimization in large-scale networks, pricing-based distributed optimization emerged as a practical alternative~\cite{gizelis2010survey}, enabling local decision-making at users and BSs through iterative price exchanges. 
This framework has been successfully adapted to a wide range of objectives: i) proportional fairness~\cite{Q_Ye_TWC,Kaiming_JSAC}, ii) max-min fairness~\cite{10254460,sun2015joint}, iii) delay-aware utility~\cite{10675431,10243579}, and iv) homogeneous $\alpha$-fairness~\cite{9738455,allybokus2018multi}. 
However, existing pricing-based optimization methods assume identical $\alpha$ values across users, limiting their flexibility in heterogeneous demands.

\subsection{Challenges and Contributions}

\paragraph*{Challenge: homogeneous fairness criterion}
While $\alpha$-fairness provides a spectrum of trade-offs---from throughput maximization ($\alpha = 0$), to proportional fairness ($\alpha = 1$), to max-min fairness ($\alpha \to \infty$)---the homogeneous application of a single $\alpha$ to all users fails to capture the inherent diversity of modern networks. In heterogeneous networks (HetNets), applications exhibit distinct performance sensitivities:
\begin{itemize}
\item MMF~\cite{sun2015joint, 10254460} ($\alpha > 4.0$) offers robustness to worst-case users but often degrades overall efficiency.
\item Delay-aware utility~\cite{10675431,10243579} ($\alpha \approx 2.0$) emphasizes latency reduction at the cost of long-term fairness.
\item PF~\cite{Q_Ye_TWC,Kaiming_JSAC} ($\alpha \rightarrow 1.0$) balances fairness and efficiency, but may fall short under mixed-priority traffic.
\item Throughput-centric strategies~\cite{9738455} ($\alpha < 1.0$) boost system throughput but risk excluding disadvantaged users.
\end{itemize}
These limitations motivate the need for a more flexible, user-aware fairness formulation.

\paragraph*{Research question}
To address the shortcomings of homogeneous fairness in UARA, we raise the following question:
\begin{mdframed}[outerlinecolor=black,outerlinewidth=1pt,linecolor=cccolor,middlelinewidth=1pt,roundcorner=0pt]
  \begin{center}
    \textit{
    How can we design a \textbf{heterogeneous} fairness criterion and the corresponding joint optimization strategy for user association and resource allocation?
    }
  \end{center}
\end{mdframed}

We answer this question by introducing a heterogeneous $\alpha$-fairness (HAF) framework, wherein each user is assigned an individual $\alpha$ value based on their QoS requirements. This allows for adaptive, context-aware trade-offs:
\begin{itemize}
\item Users with strict latency constraints (e.g., real-time control) are assigned higher $\alpha$ values.
\item Users focused on throughput (e.g., bulk transfers) are assigned lower $\alpha$ values.
\end{itemize}


\paragraph*{Our findings}

In order to address the research question and challenges we raised in the previous subsection, we propose a pricing-based framework to jointly optimize UARA for heterogeneous fairness index.
Our salient contributions are three-fold:
\begin{itemize}
    \item We propose a generalized version of the $\alpha$-fairness objective function, where the $\alpha$ value for each user is heterogeneous. 
    The proposed objective function enables us to optimize UARA for various priorities of users. 
    \item We propose a distributed pricing-based optimization method for HAF optimization inspired by Lagrangian duality. In our theoretical analysis, we show the convergence and optimality of the proposed method.
    \item In numerical results, we demonstrate group-wise network utility evaluation for various metrics, \eg, PF metric, sum-rate, latency, min-rate. The results show the proposed method can potentially manage the priority of the users by assigning different values of $\alpha$ to users. 
\end{itemize}

\begin{table}[tb]
    \caption{The decision function $f_1$ and price-updating function $f_2$ of the pricing-based UARA methods. }
    \centering
    \adjustbox{width=1\linewidth}{
    \begin{tabular}{lcc}
        \toprule
         Objective & $f_1(\gamma_{ij}, \mu_j)$ & $f_2(\{\gamma_{ij}|i\in\mathcal{I}_j\}, \mu_j)$ \\ 
        \midrule
         Sum-rate & $\gamma_{ij}$ & - \\ 
         PF~\cite{Q_Ye_TWC} & $\mu_j\gamma_{ij}$ & $\mu_j - \eta(e^{\mu_j-1} - |\mathcal{I}_j|)$ \\ 
         $\alpha$-fairness~\cite{9738455} & $\mu_j\gamma_{ij}^{\frac{1-\alpha}{\alpha}}$ & $ \mu_j - \eta\left(- \left(\frac{1-\alpha}{\alpha}\mu_j\right)^{\frac{1}{\alpha-1}} +\sum_{i\in\mathcal{I}_j}\gamma_{ij}^{\frac{1-\alpha}{\alpha}}\right)$ \\ 
         Delay~\cite{10675431} & $\mu_j/\sqrt{\gamma_{ij}}$ & $ \mu_j - \eta\left(\frac{1}{2}\mu_j+\sum_{i\in\mathcal{I}_j}1/\sqrt{\gamma_{ij}}\right)$ \\ 
         \textbf{Ours} & $\gamma_{ij} / \mu_j$ & $\mu_j - \eta\left(1 - \sum_{i\in\mathcal{I}_j}\widehat{\gamma}_{ij}\mu_j^{-\frac{1}{\alpha_i}}\right)$ \\ 
        \bottomrule
        \multicolumn{3}{l}{PF: proportional fairness}
    \end{tabular}
    }
    \label{tab:pricing_comparison}
\end{table}

\subsection{Preliminaries: Pricing-Based Optimization}

A representative approach for network utility maximization is the \textbf{pricing method}~\cite{Q_Ye_TWC,Kaiming_JSAC, Jang_19_TVT,jgjang_tvt_2, 9738455,10254460, 9718580, 10675431, 10791413, 9796120}. 
The pricing-based UARA method executes a user-centric UA at user devices, where the price of each base station BS is updated via distributed optimization~\cite{gizelis2010survey}. 
Let us assume there are $I$ users and $J$ BSs in the network, where the spectral efficiency of the link between user $i$ and BS $j$ is denoted by $\gamma_{ij}$. 
Also, we denote the price of BS $j$ as $\mu_j$.  Then, the user-centric UA makes user $i$ associate with BS $j_i^*$, where
\begin{equation}\label{eq:intro-pricing-UE}
    \textbf{At User $i$:}~~j_i^* = \argmax_{j=1,...,J} f_1(\gamma_{ij}, \mu_j),
\end{equation}
where the design of the function $f_1(\cdot)$ depends on the objective function. 
Let us denote the set of users associated with BS $j$ as $\mathcal{I}_j=\{i|j_i^*=j, i=1,...,N\}$.
Generally, every user $i$ can be associated with a BS to reduce communication overhead.
Then, on the BS side, the BSs locally update their pricing values $\mu_j$ via
\begin{equation}\label{eq:intro-pricing-BS}
    \textbf{At BS $j$:}~~\mu_j \leftarrow f_2(\{\gamma_{ij}|i\in\mathcal{I}_j\}, \mu_j),
\end{equation}
where $f_2$ is a price-updating function. 
As shown in \eqref{eq:intro-pricing-UE} and \eqref{eq:intro-pricing-BS}, the pricing-based user association does not require information exchange between BSs, thereby allowing \textbf{distributed optimization} of load balancing. 
As well as distributed implementation, another advantage of the pricing-based method is \textbf{simplicity}.

\paragraph*{Family of pricing-based methods}

Previously, a series of pricing-based UARA methods have been proposed based on the proportional fairness (PF) objective function~\cite{Q_Ye_TWC,Kaiming_JSAC, Fooladivanda_13}.
As shown in \cref{tab:pricing_comparison}, the decision function $f_1$ and price-updating function $f_2$ for the PF are defined by $f_1(\gamma_{ij}, \mu_j) = \mu_j\gamma_{ij}$ and $f_2(\{\gamma_{ij}|i\in\mathcal{I}_j\}, \mu_j) = \mu_j - \delta(e^{\mu_j-1} - |\mathcal{I}_j|)$, respectively.
Motivated by these works, several studies focus on the load balancing with QoS constraints~\cite{7387784}, space-terrestrial integrated networks~\cite{10681138}, energy-harvesting BSs~\cite{zhang2017energy}, fog networks~\cite{misra2021fogprime}, per-resource-block load balancing~\cite{Jang_19_TVT}, mobile edge computing~\cite{chou2021pricing}, and reliability optimization~\cite{lee2006network}.

Other than the PF objective function, there have been several works on the load balancing for max-min fairness~\cite{10254460,sun2015joint}, QoS-constrained sum-rate maximization~\cite {9718580,7387784}, latency minimization~\cite{10675431,10243579,chai2023joint}, and alpha-fairness~\cite{9738455,allybokus2018multi}. 

\subsection{Notations}

Given a matrix, $[\cdot]_{ij}$ denotes the $(i,j)$-th element of the matrix.
The lowercase and capital boldface variables (\eg, $\mathbf{x}$ and $\mathbf{X}$) denote a vector and matrix, respectively. 
The calligraphic letter (\eg, $\mathcal{X}$) denotes a set.

\begin{figure}[t]
    \centering
    \includegraphics[width=0.99\linewidth]{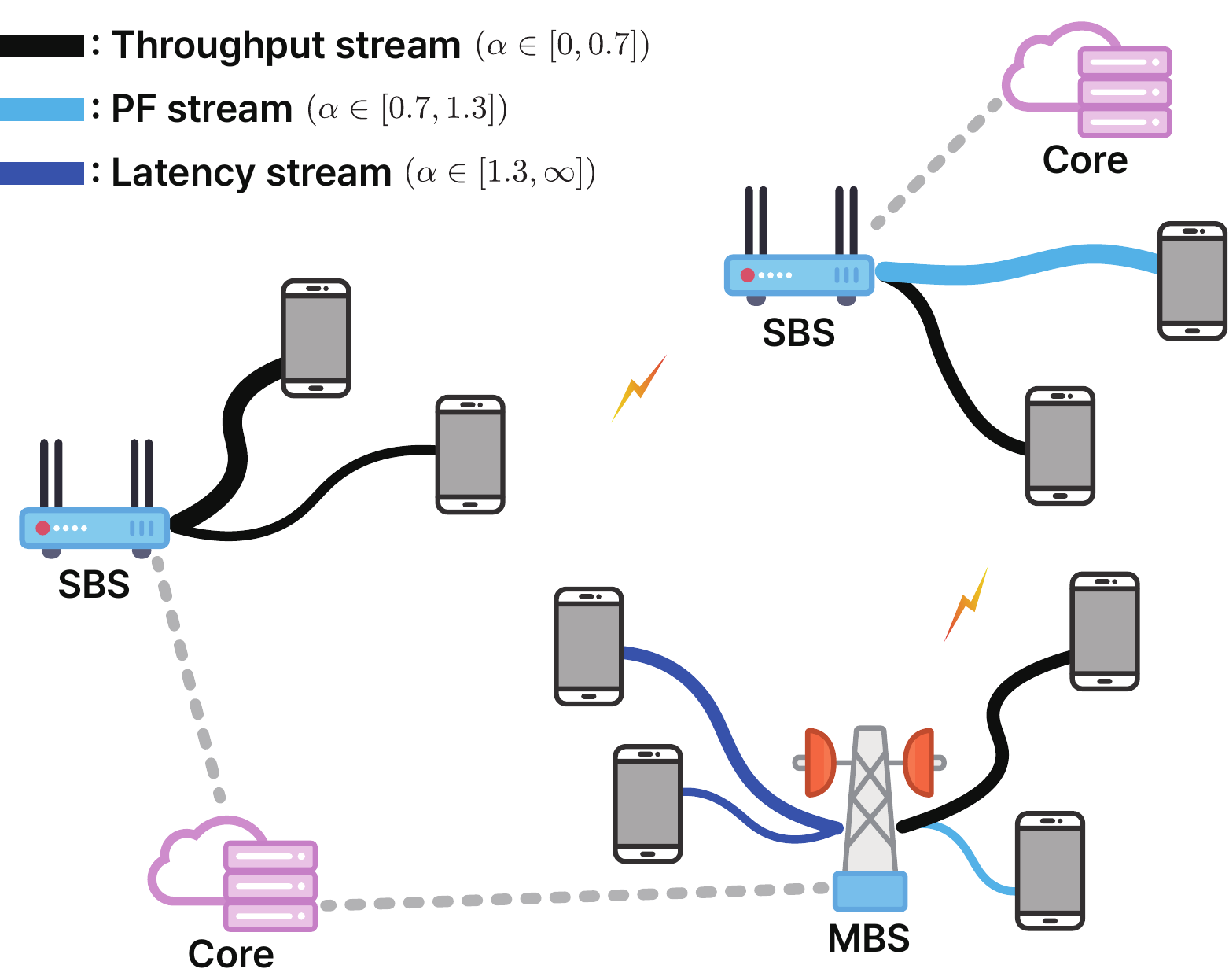}
    \caption{Illustration of the system model. The small cell BSs and macro cell BSs are co-deployed in the network. In the system model, the users have different priority levels ($\alpha$). In the low regime of $\alpha$, the users pursue throughput performance. The users with middle and high regimes of $\alpha$ pursuit PF and Latency performances, respectively. 
    }
    \label{fig:Intuition-GAF}
\end{figure}

\section{System Model and Problem Formulation}

We consider a \textit{downlink} heterogeneous network (HetNet) comprising multiple BSs and users with diverse service priorities, as illustrated in \cref{fig:Intuition-GAF}.
In the system model, there are $J$ BSs and $I$ users, where macro cell and small cell base stations are co-deployed, \eg, the third-generation project partnership (3GPP) small cell scenario 1~\cite{3gpp_36_872}.
In the remainder of the paper, we denote the index sets of the $J$ BSs and $I$ users as $\mathcal{J}=\{1,\ldots,J\}$ and $\mathcal{I}=\{1,\ldots,I\}$, respectively.
Also, we note that each of the BSs and users is equipped with a single antenna.
The backhaul link from the core network to the BSs is assumed to be nearly unlimited, \ie, fiber access in \cite{3gpp_36_932}.

In this work, we consider a frequency-division multiplexing (FDM)-based RA model, where each BS has a fixed total bandwidth, partitioned into fine-grained orthogonal resource blocks.
These frequency resource blocks are \textit{orthogonally} assigned to users associated with the BS, ensuring that no two users simultaneously occupy the same frequency resource of a BS. 

As discussed in a previous study~\cite{Q_Ye_TWC}, the many-to-many UA has more flexibility and the problem is easy to solve; however, it induces substantial communication overhead between BSs. 
Thus, for the practicality of the implementation, we assume each user can be associated with up to one BS, \ie, unique BS association.


\subsection{Communication Model}

Let us define the channel gain between BS $j\in\mathcal{J}$ and user $i\in\mathcal{I}$ at the $t$-th time slot as $h_{ij}$. 
Then, the received signal $r_{ij}$ at the user $i$ associated with the BS $j$ is denoted by 
\begin{equation}\label{eq:channel_model}
    r_{ij} = \underbrace{h_{ij} s_{j}}_{\text{signal}} +\underbrace{\sum_{k\in\mathcal{J}\setminus\{j\}} h_{ik}s_k}_{\text{interference}} + \underbrace{n_i}_{\text{noise}},
\end{equation}
where $s_{j}$ and $n_i\sim\mathcal{CN}(0,N_0)$ denote the symbol transmitted by BS $j$ to its associated user and the additive white Gaussian noise (AWGN) at user $i$, respectively.
We note that the transmitted symbol $s_j$ satisfies $\mathbb{E}[|s_j^2|] = P_j$, where $P_j$ denotes the transmission power of BS $j$. 

\paragraph*{Spectral efficiency model}
With the channel model in \eqref{eq:channel_model}, the signal-to-interference-plus-noise-ratio (SINR) of the signal transmitted from BS $j$ to user $i$ is denoted as 
\begin{equation}\label{eq:SINR}
\begin{split}
    \mathrm{SINR}_{ij} &= \frac{|h_{ij}|^2\mathbb{E}[|s_j|^2]}{\sum_{k\in\mathcal{J}\setminus\{j\}}|h_{ik}|^2\mathbb{E}[|s_k|^2]  + N_0}\\
    &= \frac{|h_{ij}|^2P_j}{\sum_{k\in\mathcal{J}\setminus\{j\}}|h_{ik}|^2P_k  + N_0}.
\end{split}
\end{equation}
Then, the spectral efficiency between BS $j$ and user $i$ is represented by
\begin{equation}
    \gamma_{ij} = \log_2(1+\mathrm{SINR}_{ij}).
\end{equation}

As depicted in \cref{fig:Intuition-GAF}, each BS can service multiple users in parallel by splitting the frequency bands, and then it allocates the split bands to the associated users, whereas each of the users can be served by up to one BS in parallel. 

\paragraph*{UA variable}
To indicate the UA of the system model, we define a binary variable $x_{ij}$ as follows:
\begin{equation}\label{eq:constraint_X}
    x_{ij} = 
    \begin{cases}
        1, \text{if user $i$ is served by BS $j$,}\\
        0, \text{otherwise.}
    \end{cases}
\end{equation}
In the later part, the augmented matrix $\mathbf{X}\in\{0,1\}^{I\times J}$ represents the all the variables $x_{ij}$ for simplicity of the notation, where $[\mathbf{X}]_{i,j} = x_{ij}$.
Furthermore, we assume that each user is associated with only one BS by using $x_{ij}$ as an indicator, which makes the problem combinatorial in nature. 
This unique BS association assumption significantly increases the complexity, especially because the UA problem should be solved in conjunction with the RA problem, as the optimal RA depends on which BS serves each user and vice versa. 
Despite the computational burden, we adopt this approach because enabling users to simultaneously associate with multiple BSs, while potentially improving theoretical performance, would introduce considerable \textit{system overhead} and implementation challenges. Thus, from a practical standpoint, the unique BS association assumption is more sensible than its multiple association counterpart.
Hence, the unique BS association assumption constraints the variable $\mathbf{X}$ by 
\begin{equation}\label{eq:constraint_X_2}
    \begin{cases}
        x_{ij} \in \{0,1\}, & \forall i\in\mathcal{I}, j\in\mathcal{J},\\
        \sum_{k\in\mathcal{J}} x_{ik} \le 1, & \forall i\in\mathcal{I}.
    \end{cases}
\end{equation}

\paragraph*{RA variable}

Let $y_{ij} \in [0,1]$ denote the fraction of the total bandwidth (\ie, the proportion of frequency resource blocks) at BS $j$ allocated to user $i$. This variable captures the share of spectrum resources assigned to each user and serves as the continuous-valued RA variable in our model.
To ensure orthogonal frequency allocation and preserve spectral exclusivity among users, we impose the following constraints:
\begin{equation}\label{eq:constraint_Y}
    \begin{cases}
        y_{ij} \in [0,1], & \forall i \in \mathcal{I},\ j \in \mathcal{J},\\
        \sum_{i \in \mathcal{I}} y_{ij} \le 1, & \forall j \in \mathcal{J}.
    \end{cases}
\end{equation}
The first condition ensures that the allocated bandwidth to any user remains within physical limits, while the second condition guarantees that the aggregate allocation across all users at a given BS does not exceed its available frequency resource. 

Analogous to the UA matrix \(\mathbf{X} \in \{0,1\}^{I \times J}\), we define the resource allocation matrix \(\mathbf{Y} \in [0,1]^{I \times J}\), where each element is given by $[\mathbf{Y}]_{i,j} =y_{ij}$. 
With these variables, the achievable rate between user $i$ and BS $j$ is modeled as $\gamma_{ij} x_{ij} y_{ij}$, which captures the bandwidth-proportional capacity under the current UA and RA decisions.


\subsection{Problem Formulation}

\paragraph*{Conventional $\alpha$-fairness}

Before presenting the HAF objective function, we review the conventional $\alpha$-fairness objective function.
The $\alpha$-fairness function is represented by 
\begin{equation}\label{eq:caf}
\sum_{i\in\mathcal{I}}\frac{\left(\sum_{j\in\mathcal{J}} \gamma_{ij}x_{ij}y_{ij}\right)^{1-\alpha}}{1-\alpha},
\end{equation}
where the parameter $\alpha\in[0,1)\cup(1,\infty)$ adjusts the weight between the fairness and efficiency of the users.
The inner summation represents the total rate of user $i$, and the outer transformation applies the $\alpha$-fair utility function, which prioritizes fairness as $\alpha$ increases.
For example, if $\alpha\rightarrow\infty$, the objective function represents the max-min fairness, \ie, $\min_{i\in\mathcal{I}}\sum_{j\in\mathcal{J}} \gamma_{ij}x_{ij}y_{ij}$. 
On the other hand, if $\alpha=0$, the $\alpha$-fairness works as sum-rate, \ie, $\sum_{i\in\mathcal{I}}\sum_{j\in\mathcal{J}}\gamma_{ij}x_{ij}y_{ij}$.

\paragraph*{Heterogeneous $\alpha$-fairness}
In this paper, we aim to control the user-wise tradeoff between the efficiency and fairness of the networks. 
As depicted in~\cref{fig:Intuition-GAF}, the users in the network request different types of streams, \eg, throughput-prioritized stream, PF-prioritized stream, and latency-prioritized stream. 
To reflect user-specific service requirements, we extend the conventional $\alpha$-fairness model to a heterogeneous formulation, where each user $i$ is assigned an individual $\alpha_i$ that governs their fairness-efficiency tradeoff, \ie,
\begin{equation}\label{eq:haf}
\sum_{i\in\mathcal{I}}\frac{\left(\sum_{j\in\mathcal{J}} \gamma_{ij}x_{ij}y_{ij}\right)^{1-\alpha_i}}{1-\alpha_i}.
\end{equation}
Different from the $\alpha$-fairness objective function~\cite{9738455}, the users have different $\alpha_i$ values, thereby enabling us to use an advanced strategy to control the user-wise tradeoff between the efficiency and fairness of the networks, \ie, flexible radio resource management for \textbf{user-wise priority}.
For example, the group of users with $\alpha\in[0,0.7]$ focuses on the throughput performance, whereas the group of users with higher $\alpha$ values pursues the fairness or delay of the services. 

\paragraph*{Problem formulation}
By integrating the objective function in Problem \ref{eq:P1} and the constraints \eqref{eq:constraint_X_2}-\eqref{eq:constraint_Y}, we formulate the joint UARA problem for the HAF maximization as
\begin{problem} \label{eq:P1}
    \begin{alignat}{3}
        \text{\ref{eq:P1}: } & & \max_{\mathbf{X},\mathbf{Y}} ~~ & \sum_{i\in\mathcal{I}}\frac{\left(\sum_{j\in\mathcal{J}} \gamma_{ij}x_{ij}y_{ij}\right)^{1-\alpha_i}}{1-\alpha_i}  \label{P1:obj} \\
        & & \text{s.t.} ~~ &  \sum_{i\in\mathcal{I}}y_{ij} \le 1 \\ 
        & & & y_{ij}\in[0,1] \\ 
        & & & \sum_{j\in\mathcal{J}}x_{ij} \le 1 \\ 
        & & & x_{ij}\in\{0,1\}.
    \end{alignat}
\end{problem}

This problem is combinatorial in nature due to binary UA variables, and the coupling between $\mathbf{X}$ and $\mathbf{Y}$ further increases the complexity, making the problem NP-hard.

\section{Proposed Lagrangian-Duality-Based Approach}


In this section, we propose an optimization algorithm for Problem \ref{eq:P1} as a form of the pricing-based approach (see \eqref{eq:intro-pricing-UE} and \eqref{eq:intro-pricing-BS}).
We begin by solving the RA subproblem for a fixed UA variable.
Although the RA solution does not admit a closed-form expression, we derive an efficient iterative approach. We then substitute the RA solution into the original problem, reformulating it as a UA optimization problem. 
This reformulated problem is tackled using Lagrangian duality, where we introduce a slack variable to ensure tractability.


\subsection{RA Optimization}

By fixing the UA variable $\mathbf{X}$, we have the following UA problem:
\begin{problem} \label{eq:P2}
    \begin{alignat}{3}
        \text{\ref{eq:P2}: } & & \max_{\mathbf{Y}} ~~ & \sum_{i\in\mathcal{I}}\frac{\left(\sum_{j\in\mathcal{J}} \gamma_{ij}x_{ij}y_{ij}\right)^{1-\alpha_i}}{1-\alpha_i}  \label{P2:obj} \\
        & & \text{s.t.} ~~ &  \sum_{i\in\mathcal{I}}y_{ij} \le 1 \\ 
        & & & y_{ij}\in[0,1].
    \end{alignat}
\end{problem}
Because the variable $\mathbf{X}$ is binary, we can rewrite the objective function of Problem \ref{eq:P2} as 
\begin{equation}
\begin{split}
    \sum_{i\in\mathcal{I}}\frac{\left(\sum_{j\in\mathcal{J}} \gamma_{ij}x_{ij}y_{ij}\right)^{1-\alpha_i}}{1-\alpha_i} \\ 
    = 
    \sum_{i\in\mathcal{I}}\sum_{j\in\mathcal{J}}\frac{\left( \gamma_{ij}y_{ij}\right)^{1-\alpha_i}}{1-\alpha_i}x_{ij}.
\end{split}
\end{equation}
Since each user's association is fixed, the original RA problem naturally decomposes across BSs. 
That is, each BS optimizes the allocation of its local bandwidth among its associated users, leading to per-BS subproblems.
By reformulating the objective function of Problem \ref{eq:P1}, we can decompose Problem \ref{eq:P2} into subproblems for BS $j$ as 
\begin{problem} \label{eq:P3}
    \begin{alignat}{3}
        \text{\ref{eq:P3}: } & & \max_{\mathbf{Y}} ~~ & \sum_{i\in\mathcal{I}_j}\frac{\left(\gamma_{ij}y_{ij}\right)^{1-\alpha_i}}{1-\alpha_i}  \label{P3:obj} \\
        & & \text{s.t.} ~~ &  \sum_{i\in\mathcal{I}_j}y_{ij} \le 1 \label{eq:P3_b}\\ 
        & & & y_{ij} \ge 0, \forall i\in\mathcal{I}_j.\label{eq:P3_c}
    \end{alignat}
\end{problem}

\begin{mdframed}[outerlinecolor=black,outerlinewidth=.5pt,linecolor=cccolor,middlelinewidth=1pt,roundcorner=0pt,  innertopmargin=0pt, innerbottommargin=3pt, innerleftmargin=8pt, innerrightmargin=8pt]
\begin{lemma}\label{lemma:KKT}
    The global optimal solution of Problem \ref{eq:P3} can be obtained by finding a non-negative $\lambda_j$ subject to 
    \begin{equation}\label{eq:optimal_Y_RA}
               \sum_{i\in\mathcal{I}} \lambda_j^{-\frac{1}{\alpha_i}}\hat{\gamma_{ij}}x_{ij} = 1,
    \end{equation}
    where $\hat{\gamma}_{ij}=\gamma_{ij}^{\frac{1}{\alpha_i}-1}$.
    The proof of this lemma can be found in Appendix \ref{sec:appendix_KKT_RA}.
\end{lemma}
\end{mdframed}

In Lemma \ref{lemma:KKT}, we obtain the condition of the optimal solution $\mathbf{Y}$ via the Karush-Kuhn-Tucker (KKT) analysis.
The optimal bandwidth allocation for BS $j$ can be derived by solving a unique $\lambda_j$ that satisfies the condition in \eqref{eq:optimal_Y_RA}. 
We note that the term $\widehat{\gamma}_{ij}=\gamma_{ij}^{(1-\alpha_i)/\alpha_i}$ denotes the fairness-adjusted spectral efficiency, which governs how much bandwidth each user should receive under the heterogeneous $\alpha_i$ weights.


\begin{algorithm}[t]
\small 
  \Function{Get\_RA($\gamma$,X,step=1e3,iters=12)}{
   \Comment{X is the UA variable matrix, (I, J)}
   \Comment{$\gamma$ is the spectral efficiency matrix, (I, J)}
   \Comment{step is the initial step size of the 1-d optimization}
   \Comment{iters is the number of RA optimization}
   \For{$j$ from 1 to $J$ (distributed)}{
    \var{
    step\_size$\leftarrow$step
    }\\
    \var{
    $\lambda_j\leftarrow 0.0$
    }\\
    \For{$k$ from 1 to iters}{
        \For{$l$ from 1 to 10}{
            \var{$\lambda_j\leftarrow \lambda_j+$step\_size} \\
            \If{$1 > \sum_{i\in\mathcal{I}}\gamma_{ij}^\frac{1-\alpha_i}{\alpha_i}\lambda_j^{-\frac{1}{\alpha_i}}x_{ij}$}{
                \var{$\lambda_j = \lambda_j -$step\_size}\\
                step\_size$\leftarrow$step\_size / 10.0
            }
        }
        \var{$\lambda_j\leftarrow\lambda_j+$step\_size}\\
    }
   }
   \var{Get $\mathbf{Y}$ by \eqref{eq:optimal_Y_RA}} \\ 
    \Return{$\mathbf{Y}$}
  }
    \caption{Resource Allocation for HAF}
      \label{algo:ra_algorithm}
\end{algorithm}


\paragraph*{Uniqueness of the solution}

To show the uniqueness of the solution $\lambda_j$ satisfying the KKT condition, we first focus on the function $\sum_{i\in\mathcal{I}} \lambda_j^{-\frac{1}{\alpha_i}} \gamma_{ij}^{\frac{1}{\alpha_i}-1}x_{ij}$. 
We note that $\alpha_i>0$ for all $i\in\mathcal{I}$, and the function $\sum_{i\in\mathcal{I}} \lambda_j^{-\frac{1}{\alpha_i}} \gamma_{ij}^{\frac{1}{\alpha_i}-1}x_{ij}$ is a continuous function.
Then, because $\lim_{\lambda\rightarrow0^+}\sum_{i\in\mathcal{I}} \lambda_j^{-\frac{1}{\alpha_i}} \gamma_{ij}^{\frac{1}{\alpha_i}-1}x_{ij} = \infty$, and since $\lim_{\lambda\rightarrow\infty}\sum_{i\in\mathcal{I}} \lambda_j^{-\frac{1}{\alpha_i}} \gamma_{ij}^{\frac{1}{\alpha_i}-1}x_{ij} = 0$, there exists at least one solution from the intermediate value theorem. 
Also, because the function $\sum_{i\in\mathcal{I}} \lambda_j^{-\frac{1}{\alpha_i}} \gamma_{ij}^{\frac{1}{\alpha_i}-1}x_{ij}$ is monotonically decreasing w.r.t. $\lambda_j$, there exists a unique $\lambda_j$ satisfying the KKT condition in \eqref{eq:appendix_lambda_2}.
Hence, by doing 1-dimensional research w.r.t. $\lambda_j$ in Algorithm~\ref{algo:ra_algorithm}, we can find the optimal solution of Problem \ref{eq:P3}.

\subsection{UA Optimization}

We now turn to optimizing the UA variable $\mathbf{X}$, given the RA solution characterized by $\lambda_j$. 
For brevity of the notation, we define an augmented vector of the $\lambda_j$ as $\Lambda=[\lambda_1, \ldots,\lambda_J]$.
In previous studies~\cite{Q_Ye_TWC,9738455,Q_Han15,Fooladivanda_13,Kaiming_JSAC}, the RA problem itself is a simple convex optimization problem; hence, it is possible to obtain a closed-form solution.
However, due to the heterogeneous $\alpha_i$ values of the users, we cannot obtain the closed-form solution. Thus, we continue the optimization of $\lambda_i$ in the next section. 

By substituting the solution in \eqref{eq:optimal_Y_RA} into Problem \ref{eq:P1}, we have the following optimization problem:
\begin{problem} \label{eq:P4}
    \begin{alignat}{3}
        \text{\ref{eq:P4}: } & & \max_{\mathbf{\Lambda},\mathbf{X}} ~~ & \sum_{j\in\mathcal{J}}\sum_{i\in\mathcal{I}}\frac{1}{{1-\alpha_i}} \widehat{\gamma}_{ij}\lambda_i^{\frac{\alpha_i-1}{\alpha_i}}x_{ij} \label{P4:obj} \\
        & & \text{s.t.} ~~ &  \sum_{j\in\mathcal{J}}x_{ij} = 1 \\ 
        & & & x_{ij}\in\{0,1\}, \forall i\in\mathcal{I},j\in\mathcal{J} \\ 
        & & & \sum_{k\in\mathcal{I}}\widehat{\gamma}_{kj}\lambda_k^{-\frac{1}{\alpha_k}}x_{kj} =1.
    \end{alignat}
\end{problem}
With the slack variable $\lambda_j$, Problem \ref{eq:P4} is still a combinatorial optimization problem requiring excessive computational complexity ($J^I$) to find the global optimal solution. 
Hence, we aim to find a sub-optimal solution via Lagrangian duality. 

\paragraph*{Duality Approach}
As a first step of our solution, we present the Lagrangian form of Problem \ref{eq:P4} as
\begin{equation}\label{eq:Lag_UA}
\begin{split}
    L_\text{UA} & = \sum_{j\in\mathcal{J}}\sum_{i\in\mathcal{I}}\frac{\widehat{\gamma}_{ij}\lambda_i^{\frac{\alpha_i-1}{\alpha_i}}x_{ij}}{{1-\alpha_i}} \\
    & + \sum_{j\in\mathcal{J}} \mu_j\left(1 - \sum_{i\in\mathcal{I}}\widehat{\gamma}_{ij}\lambda_j^{-\frac{1}{\alpha_i}}x_{ij}\right).
\end{split}
\end{equation}
Then, the Lagrangian dual function of Problem \ref{eq:P4} is represented by 
\begin{equation}
    g(\boldsymbol{\mu}) = \max_{\Lambda, \mathbf{X}} L_\text{UA}.
\end{equation}
By following the Lagrangian duality, we can obtain the sub-optimal UA $\mathbf{X}$ by finding the minimizer of $g(\boldsymbol{\mu})$, \ie,
\begin{problem} \label{eq:P5}
    \begin{alignat}{3}
        \text{\ref{eq:P5}: } & & \min_{\boldsymbol{\mu}} g(\boldsymbol{\mu}) \label{P5:obj} ~~~ \text{s.t.} ~~ &  \mu_j \ge 0.
    \end{alignat}
\end{problem}

\begin{figure}
\centering
    \includegraphics[width=0.99\linewidth]{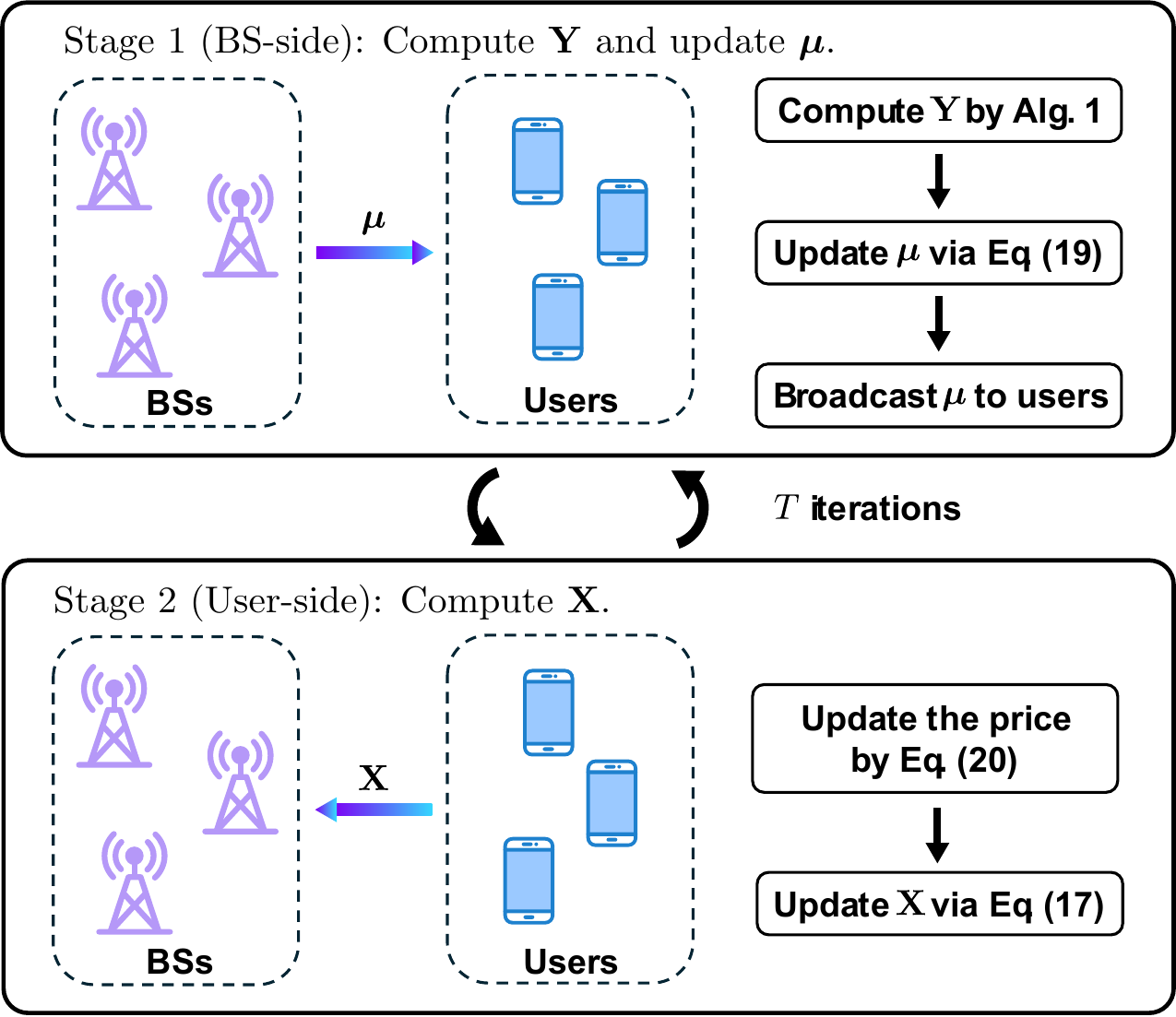}
    \caption{Illustration of the distributed optimization algorithm. }
    \label{fig:pricing-update}
\end{figure}

\paragraph*{Optimal $\Lambda$}
Here, our first focus is to find the maximize of $\Lambda$ given $\mathbf{X}$ and $\boldsymbol{\mu}$. 
Because $L_\text{UA}$ is concave w.r.t. $\Lambda$, we have 
\begin{equation}\label{eq:optimal_lambda}
    \begin{split}
        \frac{\partial L_\text{UA}}{\partial \lambda_j}& = -\sum_{i\in\mathcal{I}}\frac{\widehat{\gamma}_{ij}\lambda_j^{-\frac{1}{\alpha_i}}x_{ij}}{\alpha_i} + \mu_j\sum_{i\in\mathcal{I}}\frac{\widehat{\gamma}_{ij}\lambda_j^{-\frac{1+\alpha_i}{\alpha_i}}x_{ij}}{\alpha_i} \\ 
        & = \underbrace{\left(\sum_{i\in\mathcal{I}}\frac{\widehat{\gamma}_{ij}\lambda_j^{-\frac{1}{\alpha_i}}x_{ij}}{\alpha_i}\right)}_{>0}\left(\frac{\mu_j}{\lambda_j}-1\right) = 0\\
        & \Rightarrow \lambda_j=\mu_j.
    \end{split}
\end{equation}
By substituting $\lambda_j=\mu_j$ into \eqref{eq:Lag_UA}, we have 
\begin{equation}\label{eq:Lag_UA_2}
\begin{split}
    L_\text{UA} = \sum_{j\in\mathcal{J}}\sum_{i\in\mathcal{I}}\frac{\alpha_i\widehat{\gamma}_{ij}\mu_i^{\frac{\alpha_i-1}{\alpha_i}}x_{ij}}{{1-\alpha_i}} + \sum_{j\in\mathcal{J}} \mu_j,
\end{split}
\end{equation}
which is an affine function of $\mathbf{X}$. 

\paragraph*{Optimal $\mathbf{X}$}
Because $x_{ij}\in\{0,1\}$ and $\sum_{j\in\mathcal{J}}x_{ij} = 1$, the optimal $\mathbf{X}$ maximizing \eqref{eq:Lag_UA_2} can be obtained by finding the index $j$ with the maximum value of $\frac{\alpha_i \widehat{\gamma}_{ij}}{1-\alpha_i}\mu_j^{\frac{\alpha_i-1}{\alpha_i}}$.
From our assumption on $\alpha_i$, we have $\alpha_i\in(0,\infty)$.
Hence, the optimal $\mathbf{X}$ can be rewritten by 
\begin{equation}\label{eq:optimal_X}
    x_{ij}^* = \begin{cases}
        1, & \text{if~} j = \argmax_{k} \frac{\gamma_{ik}}{\mu_k},\\
        0, & \text{otherwise.}
    \end{cases}
\end{equation}
Substituting the optimal value of $\mathbf{X}$, we can obtain the dual function $g(\boldsymbol{\mu})$ as 
\begin{equation}\label{eq:func_g}
    g(\boldsymbol{\mu}) = \sum_{j\in\mathcal{J}}\mu_j + \sum_{i\in\mathcal{I}}\max_{j\in\mathcal{J}} \frac{\alpha_i}{1-\alpha_i}\widehat{\gamma}_{ij}\mu_j^{\frac{\alpha_i-1}{\alpha_i}}.
\end{equation}

\paragraph*{Pricing-based optimization}
Here, we aim to solve Problem \ref{eq:P5}, where the dual function is derived in \eqref{eq:func_g}. We note that the function $g(\boldsymbol{\mu})$ is a convex function because it is the maximum of the affine functions. However, the function is non-differentiable due to the $\max$ operator. 
Hence, we use a sub-gradient descent method to find the solution, where the sub-gradient descent update of the function $g$ is defined by 
\begin{equation}\label{eq:subgrad}
    \mu_j^{(t+1)}\leftarrow \mu_j^{(t)} - \eta\left(1 - \widehat{\gamma}_{ij}\mu_j^{-\frac{1}{\alpha_i}}\right),
\end{equation}
where $t$ denotes the index of the iteration and $\eta$ denotes the step size at the $t$-th iteration.

\begin{remark}[Standard form]
To align our formulation with the canonical structure of pricing-based optimization, we express our approach in the standard $f_1$ and $f_2$ form as follows:
\begin{equation}
    \begin{cases}
        f_1(\gamma_{ij},\mu_j) = \frac{\gamma_{ij}}{\mu_j}\\
        f_2(\{\gamma_{ij}|i\in\mathcal{I}_j\}, \mu_j) = \mu_j - \eta\left(1 - \widehat{\gamma}_{ij}\mu_j^{-\frac{1}{\alpha_i}}\right).
    \end{cases}
\end{equation}
\end{remark}

In \cref{fig:pricing-update}, we illustrate the iterative pricing-based optimization algorithm for the HAF objective function. 
\begin{itemize}
    \item \textbf{User side update: } Each user finds the target BS by using the broadcasted pricing value $\mu_j$. (Equation \eqref{eq:optimal_X}). 
    \item \textbf{BS side update: } From the users' decision, each BS locally updates the pricing by the sub-gradient descent in \eqref{eq:subgrad}; then, the BS broadcasts the price value $\mu_j$ to the users. 
\end{itemize}
Formal algorithm of the proposed method is given in Algorithm \ref{algo:ua_algorithm}

\begin{algorithm}[t]
\small
    \label{algo:ua_algorithm}
	\caption{Joint UARA Algorithm for HAF maximization}
    \Inp{Initial pricing variables and other parameters}{
        $T$: Total iterations\;
        $\Lambda^{(1)}$: Initial pricing variables\; 
        $\eta$: Step size of the sub-gradient descent\; 
        $\mathbf{X}$: Initial user association\;
    }
    
    \For(){each integer $t$ in $\{1,...,T\}$}{
        \Comment{Stage 1}
            $\mathbf{Y}\leftarrow\mathtt{Get\_RA}(\gamma,\mathbf{X})$\;
            $\mu_j^{(t+1)}\leftarrow \mu_j^{(t)} - \eta\left(1 - \widehat{\gamma}_{ij}\mu_j^{-\frac{1}{\alpha_i}}\right),\forall j\in\mathcal{J}$\;
        \Comment{Stage 2: Do in parallel for each $i$}
            \For(){each integer $i$ in $\mathcal{I}$}{    
                \For(){each integer $j$ in $\mathcal{J}$}{    
                    $x_{ij}^* = \begin{cases}1, & \text{if~} j = \argmax_{k} \frac{\gamma_{ik}}{\mu_k},\\0, & \text{otherwise.}\end{cases}$\;
                    }
                }
    }
\end{algorithm}

\section{Theoretical Analysis}
\label{sec:theoretical_analysis}

In this section, we provide theoretical results regarding the proposed method. 
First, we show that the proposed method converges to $\epsilon$-optimal solution of Problem \ref{eq:P5} within $\mathcal{O}(\epsilon^2)$ iterations. 
Second, we provide the optimality analysis of the proposed method. 

\subsection{Convergence Analysis}

In this analysis, we assume that the optimal value of the dual function $\min_{\boldsymbol{\mu}}g(\boldsymbol{\mu})$ is lower-bounded, \ie,~$g(\boldsymbol{\mu})>-\infty$. 
Let us denote the price $\boldsymbol{\mu}$ at the $t$-th iteration of Algorithm \ref{algo:ua_algorithm} as $\boldsymbol{\mu}^{(t)}$.
Then, we show the convergence of Algorithm \ref{algo:ua_algorithm} in Theorem \ref{thm:convergence}. 

\begin{mdframed}[outerlinecolor=black,outerlinewidth=.5pt,linecolor=cccolor,middlelinewidth=1pt,roundcorner=0pt,  innertopmargin=0pt, innerbottommargin=3pt, innerleftmargin=8pt, innerrightmargin=8pt]
\begin{theorem}
    \label{thm:convergence}
    Define the optimal solution of Problem \ref{eq:P5} as $\boldsymbol{\mu}^*$.
    Also, we further denote the sub-gradient vector in \eqref{eq:subgrad} as $\Vert\mathbf{g}_t\Vert \le G$ for all $t\in\mathbb{N}$, where $[\mathbf{g}]_i = 1 - \widehat{\gamma}_{ij}\mu_j^{-\frac{1}{\alpha_i}}$.
    Then, if $\eta = \frac{\Vert\boldsymbol{\mu}^{(1)} - \boldsymbol{\mu}^* \Vert}{G\sqrt{T}}$, the objective function of Algorithm \ref{eq:P5} converges like
    \begin{equation}
        \begin{split}
            \min_{t\in T} g(\boldsymbol{\mu}^{(t)}) - g(\boldsymbol{\mu}^*) \le \frac{G\Vert\boldsymbol{\mu}^{(1)} - \boldsymbol{\mu}^* \Vert^2}{\sqrt{T}},
        \end{split}
    \end{equation}
    where $T$ denotes the number of iterations of Algorithm \ref{algo:ua_algorithm}. 
    We provide the step-by-step proof of this theorem in Appendix \ref{appendix:convergence}.
\end{theorem}
\end{mdframed}


Theorem \ref{thm:convergence} ensures that the pricing variable $\boldsymbol{\mu}$ converges to a solution with bounded optimality gap $\epsilon$, which diminishes with the number of iterations $T$ as $\mathcal{O}(1/\sqrt{T})$. 
This supports the practical efficiency of our distributed sub-gradient method.

\subsection{Optimality Analysis}

In Problem \ref{eq:P5}, we handle the variable $\Lambda$ as a slack variable; however, in our implementation, we actually use the value of $\Lambda$ obtained from Algorithm \ref{algo:ra_algorithm}. 
In this section, we bridge the gap between the HAF objective function obtained from Algorithm \ref{algo:ua_algorithm} and its upper bound.

\begin{mdframed}[outerlinecolor=black,outerlinewidth=.5pt,linecolor=cccolor,middlelinewidth=1pt,roundcorner=0pt,  innertopmargin=0pt, innerbottommargin=3pt, innerleftmargin=8pt, innerrightmargin=8pt]
\begin{theorem}
    Let $f^*$ be the HAF obtained by implementing Algorithm \ref{algo:ua_algorithm}. 
    Denoting $f_\text{opt}$ as the global optimal solution of the problem, the gap between the obtained solution and the global optimal solution is bounded as follows:
    \label{thm:optimality}
        \begin{equation}
        \begin{split}
            f_\text{opt} - f^* & \le \sum_{j\in\mathcal{J}}(\lambda_j^*-\widehat{\lambda}_j) \\ 
            & ~~~~~ + \sum_{i\in\mathcal{I}}\sum_{j\in\mathcal{J}}\frac{\alpha_i\hat{\gamma}_{ij}x_{ij}}{1-\alpha_i} \left((\lambda_j^*)^\frac{\alpha_i-1}{\alpha_i} - (\widehat{\lambda}_j)^\frac{\alpha_i-1}{\alpha_i} \right),
        \end{split}
    \end{equation}
    where $\widehat{\Lambda}$ and $\Lambda^*$ are 
    The proof of this theorem is shown in Appendix \ref{sec:appendix_optimality}. 
\end{theorem}
\end{mdframed}


Theorem \ref{thm:optimality} characterizes the optimality gap between the algorithm's output and the global optimum in terms of the pricing variable $\Lambda$. When $\hat{\Lambda}$ approaches $\Lambda^*$, the HAF performance becomes nearly optimal, validating the efficiency of our two-stage design.

\begin{table}[t]
    \caption{Channel modeling parameters used in the simulation.}
    \label{tab:channel_modeling}
    \centering
    \begin{tabular}{cc}
    \toprule
        Parameters & Value \\
    \midrule
        Number of BSs $J$ & 6 \\ 
        Number of Users $I$ & 40 to 60 \\ 
        Bandwidth (MHz) & 20 \\ 
        Transmission power (dBm) & 23 to 36 \\ 
        Cell size (m) & 250 \\ 
        Noise power (dBm/Hz) & -174 \\ 
        Indoor probability (\%) & 50 \\ 
        Simulator & NVIDIA Sionna \\ 
    \bottomrule
    \end{tabular}
\end{table}

\section{Experimental Results}

We evaluate the proposed HAF-based UARA framework through extensive simulations under 3GPP small-cell scenarios, comparing its performance against several baseline methods across static and time-varying channels.


\paragraph*{Simulation setup}
Our simulations consider a HetNet composed of 6 BSs and 40 to 60 users.
The transmission power of the top 10\% of BSs is 33 to 36 dBm. 
For the remainder of the BSs, the transmission power is 23 to 30 dBm. 
Also, the number of the small cell clusters is assumed to be 3, where the inter-cluster interference is negligibly small compared to intra-cluster interference~\cite{R1_131635}. 
Each user's $\alpha_i$ is randomly chosen from the interval $\mathcal{A}_1=[0.4, 0.6]$, $\mathcal{A}_2=[0.7, 0.9]$, $\mathcal{A}_3=[1.8, 2.2]$, and $\mathcal{A}_4=[2.75, 3.25]$, where the ratio of choice is $\mathcal{A}_1:\mathcal{A}_2:\mathcal{A}_3:\mathcal{A}_4=0.25:0.25:0.25:0.25$ in the low fairness scenario and $\mathcal{A}_1:\mathcal{A}_2:\mathcal{A}_3:\mathcal{A}_4=0.25:0.125:0.19:0.375:0.31$ in the high fairness scenario. 
The $\alpha$-ranges represent different classes of user requirements, from highly throughput-centric ($\mathcal{A}_1$) to strongly fairness-sensitive ($\mathcal{A}_4$), reflecting heterogeneous service demands.
The detailed channel modeling parameters are listed in \cref{tab:channel_modeling}, which follows the 3GPP small cell simulation document in~\cite{3gpp_38_901}. 
In the experiments, we randomly generate 1,000 samples for each scenario and obtain average experimental results by using NVIDIA Sionna~\cite{sionna}. 

\paragraph*{Baselines}
For comparison, we consider the following baseline schemes.
\begin{itemize}
    \item \textbf{Random association (Random): } Each user picks BS association from the set $\mathcal{J}$ with uniform probability distribution. For the RA optimization, we use the proposed RA algorithm in Algorithm \ref{algo:ra_algorithm}. 
    \item \textbf{Max-SINR~\cite{3gpp_36_872}: } Each user picks a BS with the maximum SINR, \ie, $f_1(\gamma_{ij},\mu_j)=\gamma_{ij}$. We use the proposed RA algorithm for the RA optimization. 
    \item \textbf{PF~\cite{Q_Ye_TWC}: } A pricing-based UARA optimization approach that maximizes the proportional fairness. We note that this is a special case of $\alpha$-fairness with $\alpha=1$. 
    \item \textbf{$\alpha$-fairness-low (AF-Low)~\cite{9738455}: } A pricing-based UARA optimization algorithm for the $\alpha$-fairness, where $\alpha$ of the users is fixed to $0.6$. 
    \item \textbf{$\alpha$-fairness-high (AF-High)~\cite{9738455}: } A pricing-based UARA optimization algorithm for the $\alpha$-fairness, where $\alpha$ of the users is fixed to $1.6$. 
    \item \textbf{Min-Latency~\cite{10675431}: } A pricing-based UARA optimization algorithm for the latency minimization.
\end{itemize}

In addition to the above \textit{\textbf{distributed}} algorithms, we additionally implement the following \textbf{\textit{centralized}} algorithms.
\begin{itemize}
    \item \textbf{2-distance ring solution (2RS)~\cite{Y_Xia_ring_sol}: } This method finds a local optimal point of a combinatorial optimization problem. Let $c$ be the cost function, the algorithm finds $\mathbf{X}$ that satisfies $c(\mathbf{X}) \le c(\mathbf{X}')$ for all $\Vert\mathbf{X} - \mathbf{X}'\Vert_0\le 2$. 
    \item \textbf{Genetic Algorithm (GA)~\cite{GA}:} A genetic algorithm with 60 populations, 10 parents matching, mutation probability of 1\%, and a maximum of 300 generations. Since the GA for the joint UARA problem requires excessive computation time, we implement GA only for UA optimization \ie, we use the proposed RA optimization algorithm for $\mathbf{Y}$.
\end{itemize}

\begin{figure}
    \centering
    \includegraphics[width=1.0\linewidth]{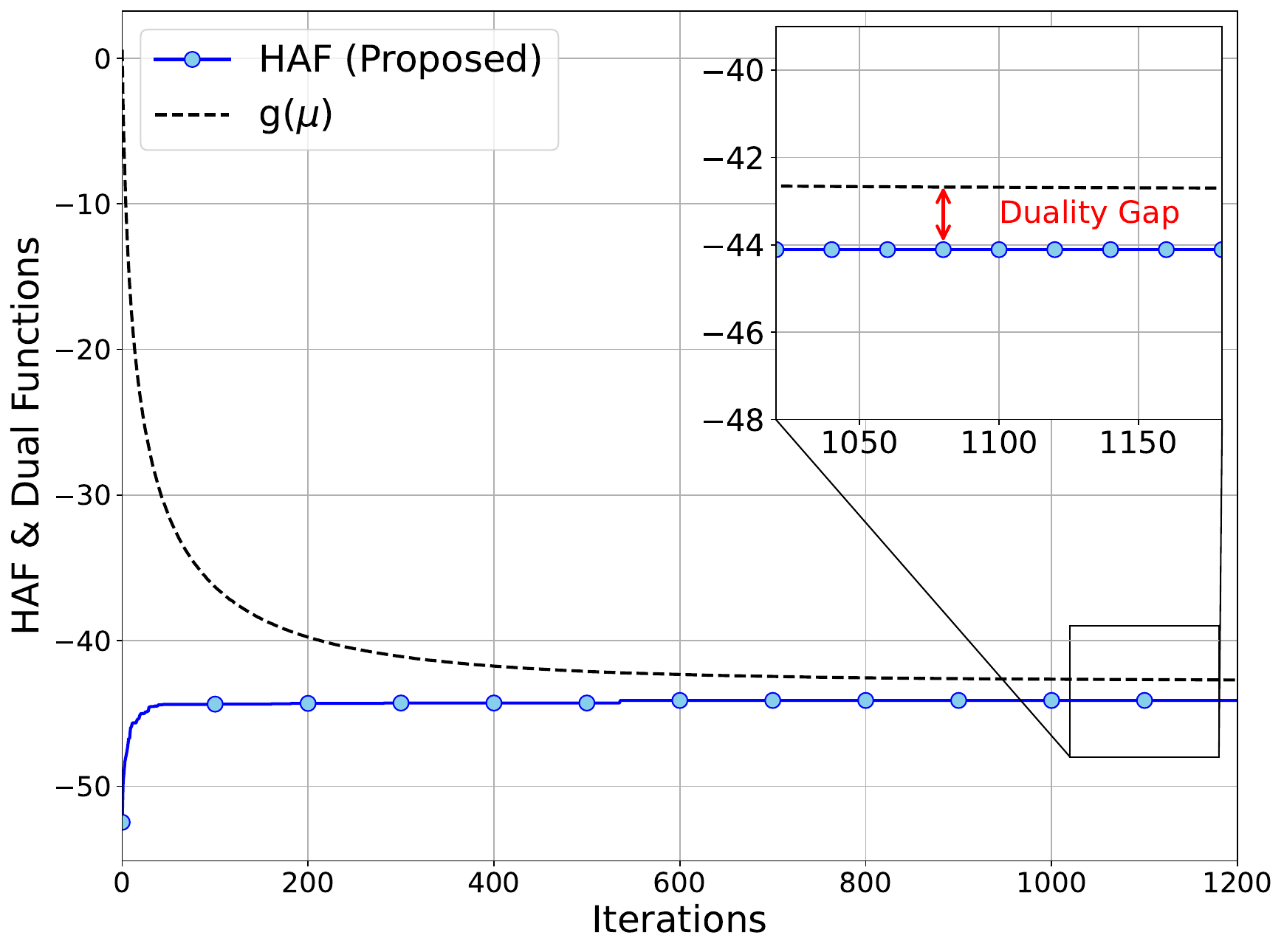}
    \caption{Illustration of the convergence of the proposed method. }
    \label{fig:convergence}
\end{figure}

\subsection{Convergence and Optimality Analysis}

In this subsection, we analyze the convergence of the proposed algorithm in Algorithm \ref{algo:ua_algorithm}. 
In \cref{fig:convergence}, we depict HAF objective function values and the dual function of the proposed scheme for each iteration to show the convergence. 
As shown in the figure, the dual function converges to the minimum point as more iterations are implemented. 
Furthermore, the HAF objective function rapidly increases in the initial stage of the algorithm.
More importantly, as discussed in Theorem \ref{thm:optimality}, the proposed method achieves the globally optimal solution if $\widehat{\Lambda}=\Lambda^*$. 
However, because $\widehat{\Lambda}\neq\Lambda^*$ in our real implementation, the \textbf{\textit{dual function is an upper bound}} of the HAF.
As depicted in the figure, the proposed method closely achieves the upper bound of the HAF.

\subsection{Fixed Channel Model}

\begin{table}[ht]
    \caption{Overall HAF and the group-wise HAF of the proposed method and the baseline methods in the \textbf{normal} scenario. We note that the GA and 2RS are the \textbf{centralized method}. Furthermore, the GA scheme requires \textbf{excessive computational complexity}. }
    \label{tab:HAF_table_normal_scenario}
    \centering
    \adjustbox{width=1\linewidth}{
    \begin{tabular}{cccccc}
         \toprule
         & HAF & HAF@$\mathcal{A}_1$ & HAF@$\mathcal{A}_2$ & HAF@$\mathcal{A}_3$ & HAF@$\mathcal{A}_4$ \\
        \midrule
            \textbf{Ours} & \textbf{70.543} & \textbf{27.588} & 58.780 & \textbf{-10.262} & \textbf{-5.563} \\
            Random & -1.490e+15 & 2.172e+00 & 1.158e+01 & -1.271e+07 & -1.490e+15 \\
            Max-SINR & 62.890 & 24.699 & 56.857 & -11.713 & -6.953 \\
            PF & 63.128 & 24.907 & \textbf{58.985} & -11.040 & -9.724 \\
            AF-Low & -114.594 & 25.426 & 57.581 & -28.130 & -169.470 \\
            AF-High & 57.274 & 23.014 & 57.291 & -12.107 & -10.924 \\
            Min-Latency & 61.181 & 23.294 & 57.589 & -11.200 & -8.502 \\
        \midrule
            2RS & 70.599 & 27.488 & 58.831 & -10.189 & -5.532 \\
            GA & 69.772 & 26.949 & 58.678 & -10.252 & -5.596 \\
         \bottomrule
         \multicolumn{6}{l}{* The best method among \textit{distributed optimization methods} is marked \textbf{bold}.}
    \end{tabular}
    }
\end{table}

\paragraph*{Low Fairness Scenario}

Here, we compare the HAF performance of the proposed method with the baseline schemes. Here, each user's $\alpha_i$ is drawn from the ratio of $\mathcal{A}_1:\mathcal{A}_2:\mathcal{A}_3:\mathcal{A}_4=0.25:0.25:0.25:0.25$.
In Table \ref{tab:HAF_table_normal_scenario}, we show the total HAF and group-wise HAF in the low-fairness scenario.
In the table, we compared the proposed method with the decentralized optimization methods (Random, Max-SINR, AF-Low, AF-High, Min-Latency) and centralized optimization methods (2RS and GA), where we represent the best of the decentralized schemes as \textbf{bold} characters.
As shown in the table, the proposed method closely achieves the HAF of the centralized optimization methods by solving the optimization problem in Problem \ref{eq:P1}. 
From this result, we show the proposed form of pricing-based optimization is more appropriate compared to the existing pricing-based methods.
Moreover, let us consider group-wise HAF performances.
The proposed method outperforms all the baselines except Group 2 ($\mathcal{A}_2$).
We note that there has been a tradeoff between the metrics because the system's radio resources are limited. 
Despite this, there is a negligible performance gap between the proposed method and the PF scheme. 

\begin{figure}
    \centering
    \includegraphics[width=0.99\linewidth]{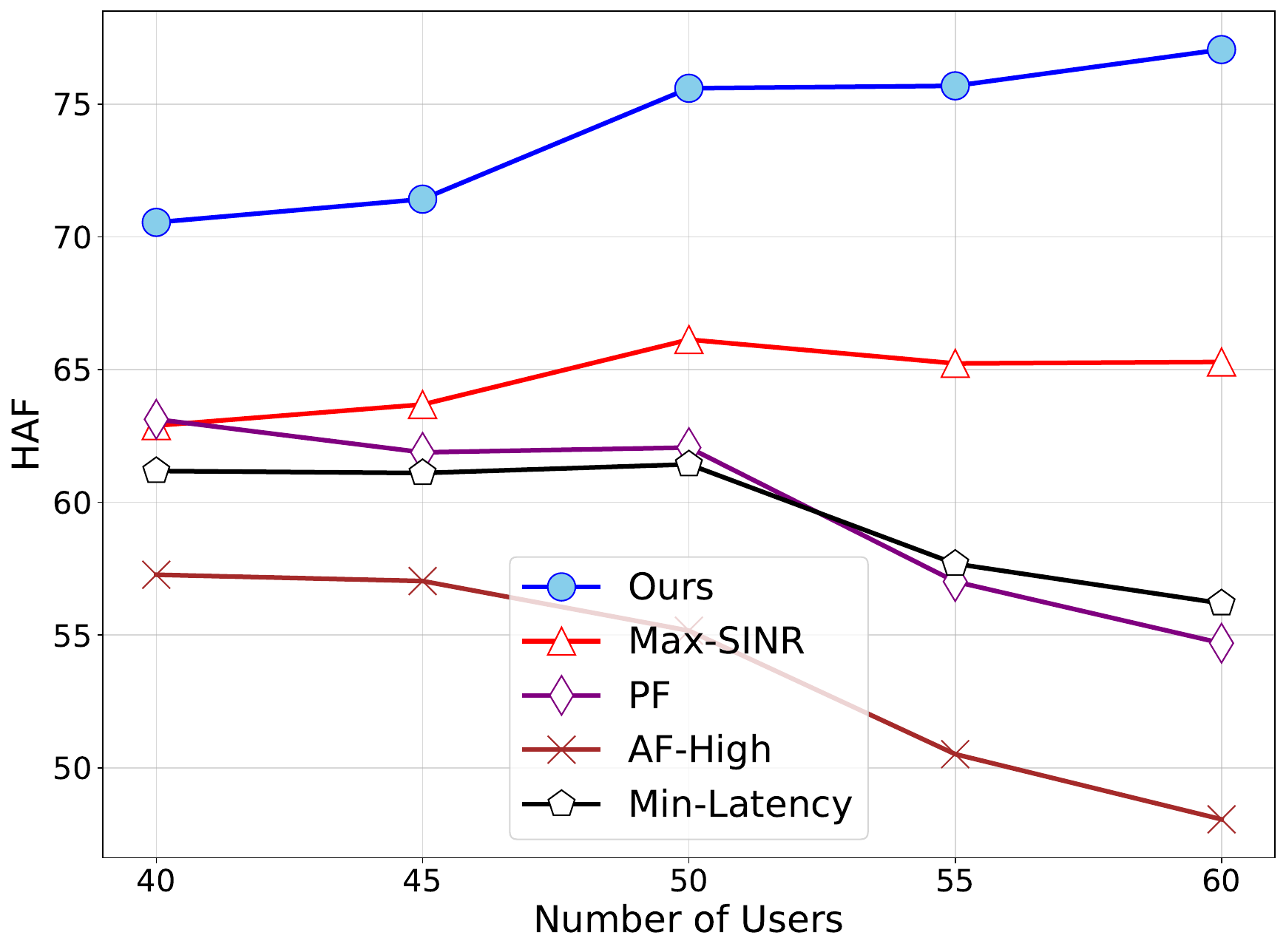}
    \caption{HAF performances of the proposed method and baseline schemes for various numbers of users. }
    \label{fig:HAF_vs_users_normal}
\end{figure}

\begin{figure}[tb]
    \centering
    \subfloat[Sum-rate ($\mathcal{A}_1$).\label{subfig:metrics_normal_40_a}]{\includegraphics[width=\if 1\doublecolumn .49 \else 0.6 \fi\linewidth]{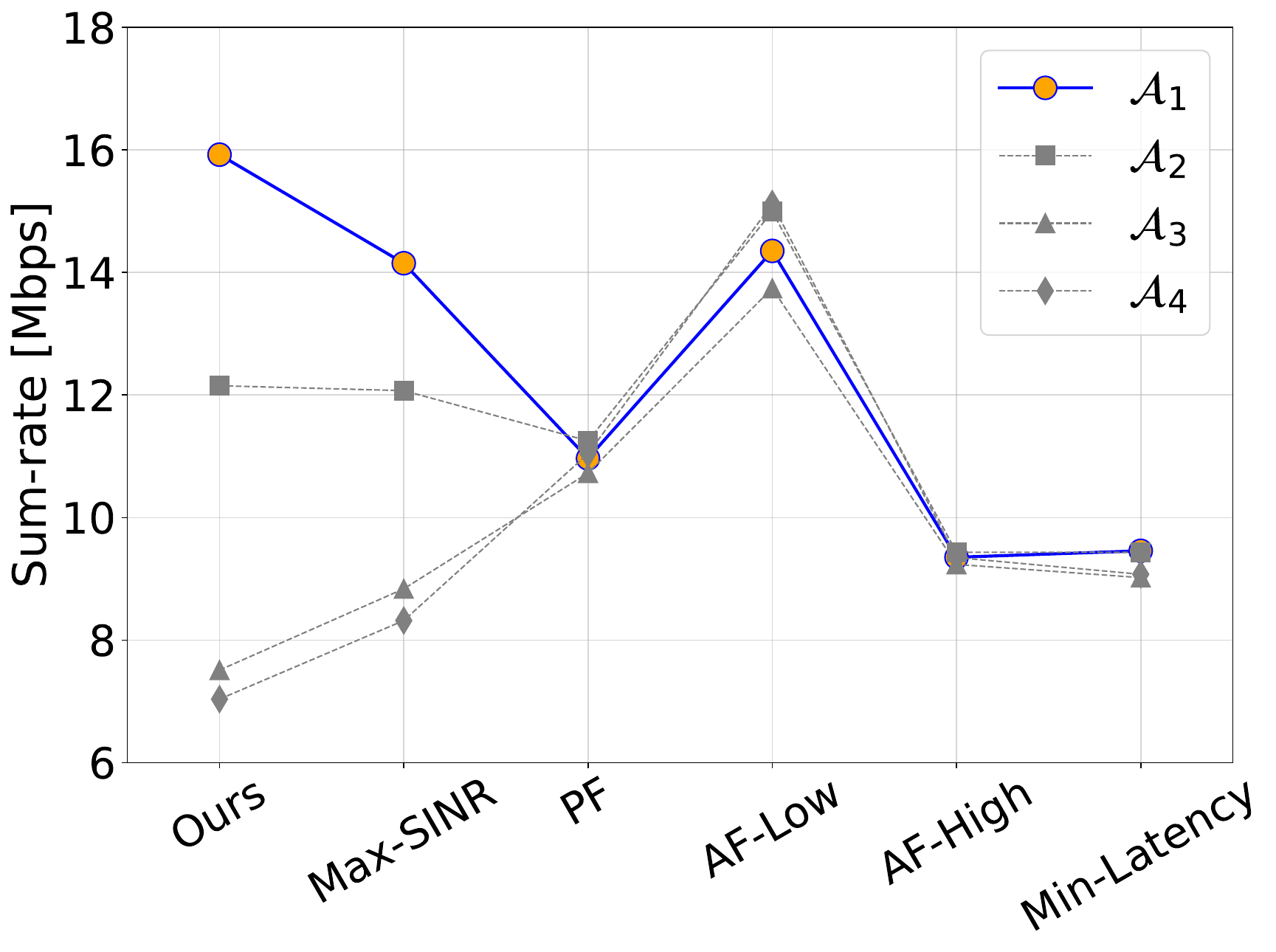}}
    \subfloat[Proportional Fairness ($\mathcal{A}_2$).\label{subfig:metrics_normal_40_b}]{\includegraphics[width=\if 1\doublecolumn .49 \else 0.6 \fi\linewidth]{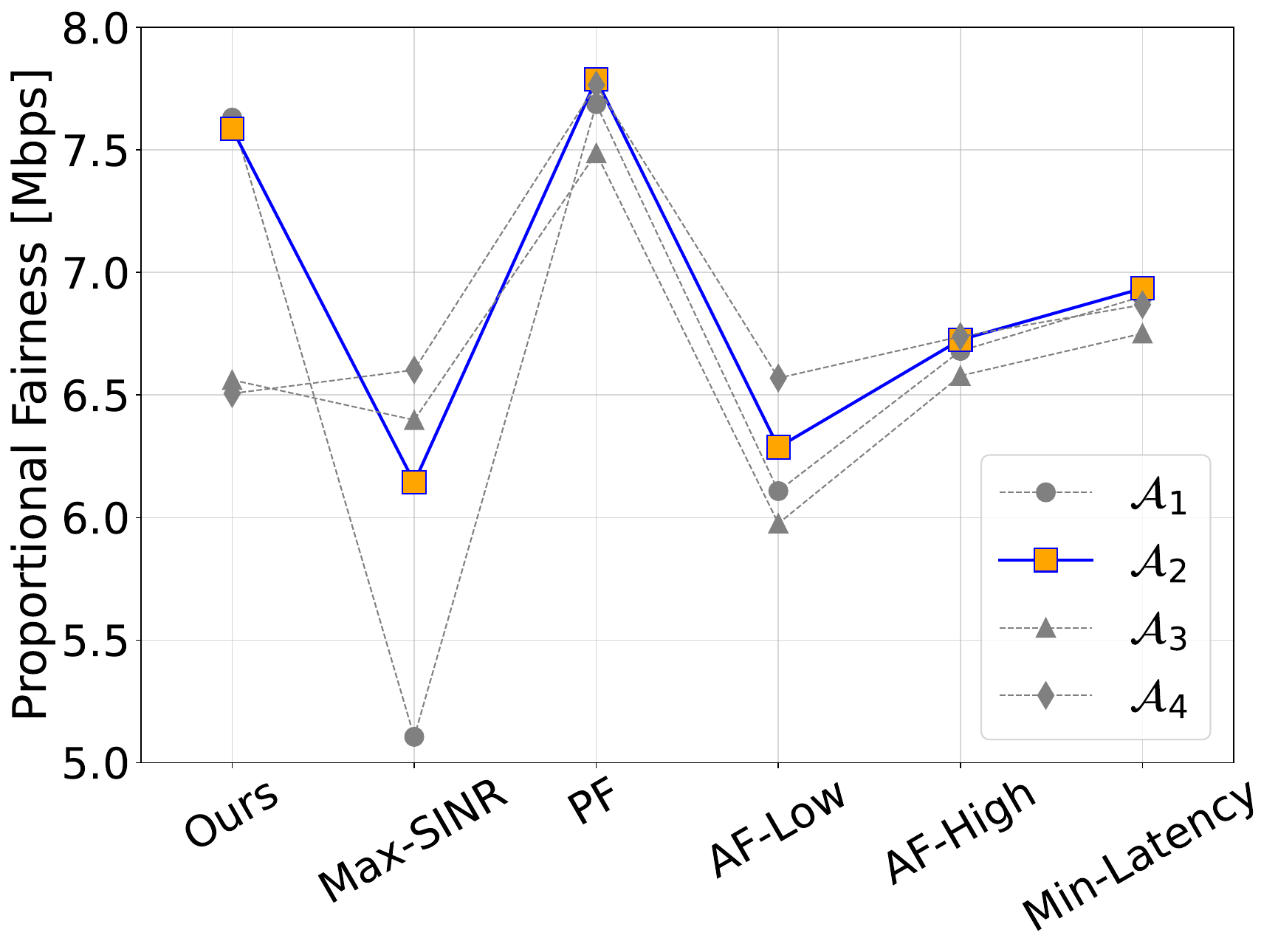}}\\
    \subfloat[Average Latency ($\mathcal{A}_3,\mathcal{A}_4$).\label{subfig:metrics_normal_40_c}]{\includegraphics[width=\if 1\doublecolumn .49 \else 0.6 \fi\linewidth]{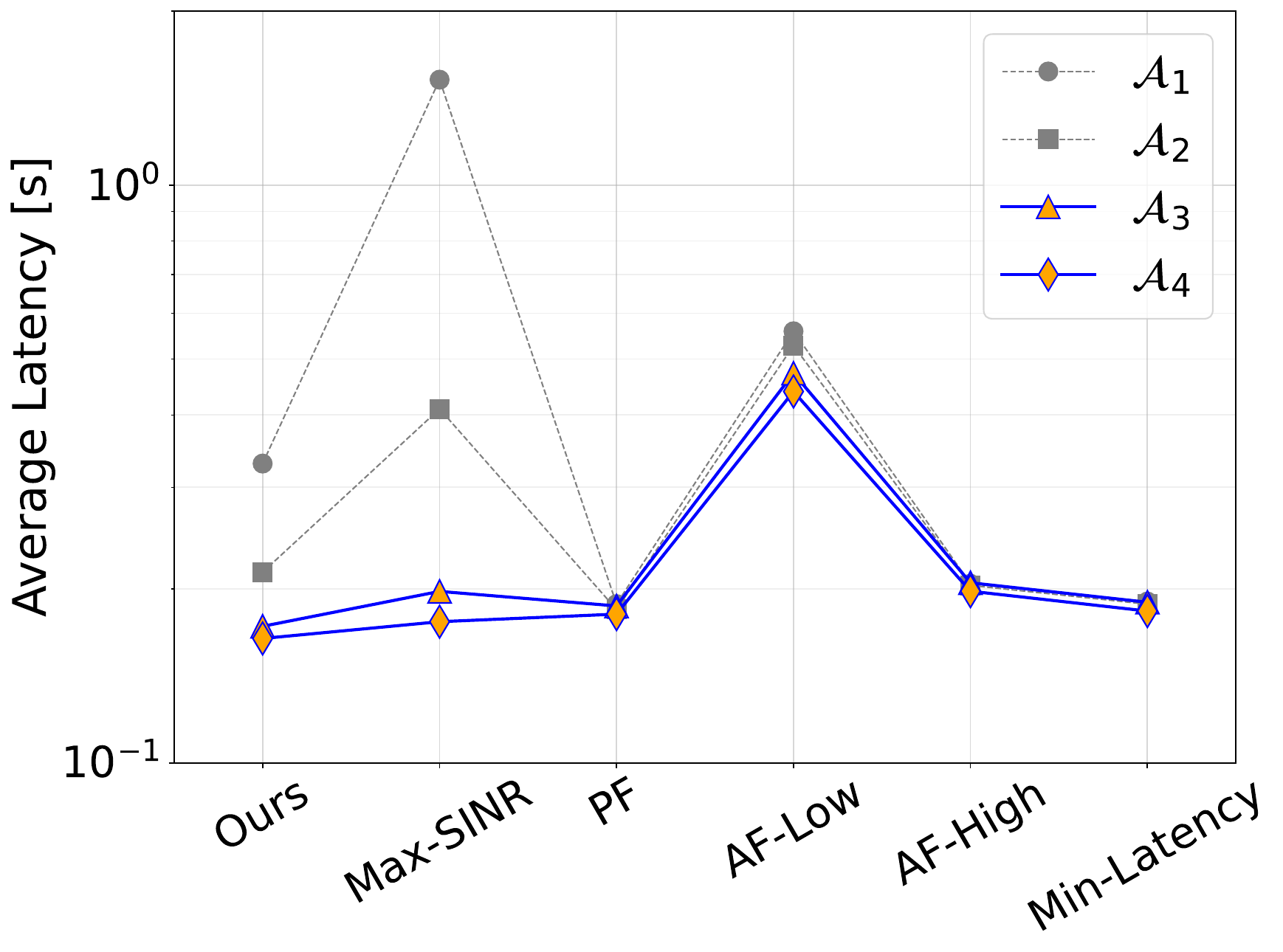}}
    \subfloat[Min-Rate ($\mathcal{A}_4$).\label{subfig:metrics_normal_40_d}]{\includegraphics[width=\if 1\doublecolumn .49 \else 0.6 \fi\linewidth]{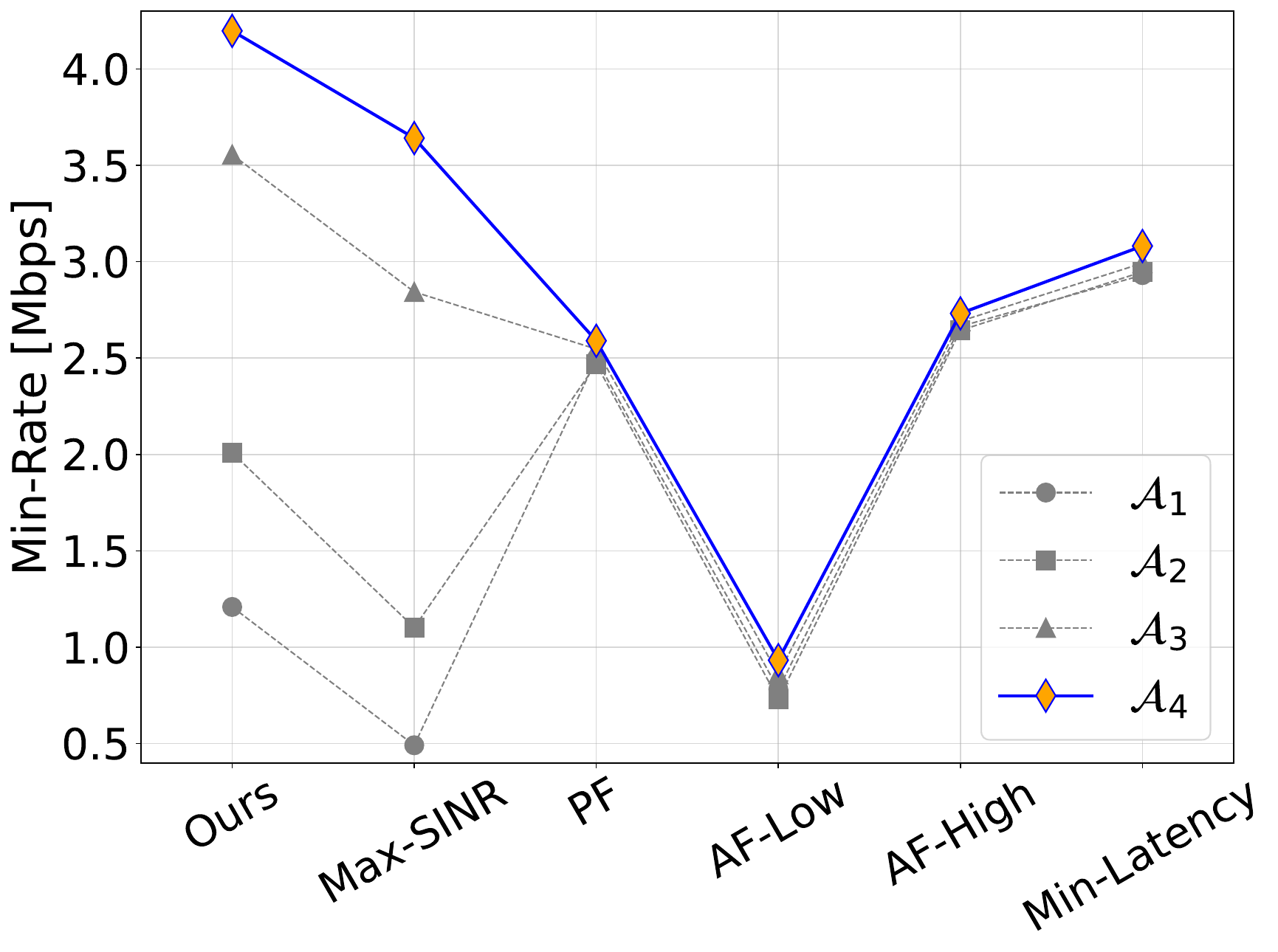}}
    
    \caption{Per-group metrics of the proposed method and baseline schemes: \protect\subref*{subfig:metrics_normal_40_a} Sum-rate, \protect\subref*{subfig:metrics_normal_40_b} Proportional fairness, \protect\subref*{subfig:metrics_normal_40_c} Average latency, and \protect\subref*{subfig:metrics_normal_40_d} Min-rate. 
    The groups corresponding to the metric are highlighted in \textcolor{blue}{solid line}, whereas the other groups are represented by \textcolor{gray}{dashed line}.
    }
    \label{fig:metrics_normal_40}
\end{figure}

In \cref{fig:HAF_vs_users_normal}, we depict the HAF performance of the proposed method and baseline schemes by varying the numbers of users from 40 to 60. 
As shown in the figure, the proposed method outperforms the baseline schemes. We note that we did not depict the Random and AF-Low schemes due the exceptionally low performance and 2RS and GA schemes due to the exceptionally high computational complexity. 
Interestingly, the proposed method's HAF increases as the number of users, whereas those of the baselines generally decreases. 
This is because the proposed method jointly optimizes the UARA with the consideration of each user's $\alpha$, which is more crucial if radio resources are scarce.

To further analyze the group-wise metrics, we depict \subref*{subfig:metrics_normal_40_a} sum-rate, \subref*{subfig:metrics_normal_40_b} proportional fairness, \subref*{subfig:metrics_normal_40_c} average latency, and \subref*{subfig:metrics_normal_40_d} min-rate of each group in \cref{fig:metrics_normal_40}. 
In \cref{subfig:metrics_normal_40_a}, since the metric is sum-rate, the target group is Group $\mathcal{A}_1$.
As shown, the proposed method outperforms the baselines, especially for the target group.
Interestingly, the difference between each group's sum-rate is large in the proposed method, whereas the other pricing-based methods have similar performance for all groups. 
This shows how the proposed method outperforms the baselines in the total HAF. 
From \cref{subfig:metrics_normal_40_b,subfig:metrics_normal_40_c,subfig:metrics_normal_40_d}, the proposed method outperforms the baseline schemes for the targeting user groups except \cref{subfig:metrics_normal_40_b}. 
For the proportional fairness metric, the PF scheme has slightly higher PF compared to the proposed method; however, the proposed method highly outperforms the PF schemes for the other metrics. 
For example in \cref{subfig:metrics_normal_40_d}, the proposed method has a 1.6x higher min-rate compared to the PF scheme. 

\begin{table}[tb]
    \caption{Overall HAF and the group-wise HAF of the proposed method and the baseline methods in the \textbf{high} scenario. We note that the GA and 2RS are the \textbf{centralized method}. Furthermore, the GA scheme requires \textbf{excessive computational complexity}. }
    \label{tab:HAF_table_high_scenario}
    \centering
    \adjustbox{width=1\linewidth}{
    \begin{tabular}{cccccc}
         \toprule
         & HAF & HAF@$\mathcal{A}_1$ & HAF@$\mathcal{A}_2$ & HAF@$\mathcal{A}_3$ & HAF@$\mathcal{A}_4$ \\
         \midrule
            \textbf{Ours} & \textbf{34.229} & \textbf{13.318} & 44.437 & \textbf{-16.285} & \textbf{-7.240} \\
            Random & -3.855e+14 & 8.119e-01 & 7.588e+00 & -1.196e+08 & -3.855e+14 \\
            Max-SINR & 27.083 & 11.929 & 43.099 & -18.644 & -9.301 \\
            PF & 28.669 & 12.222 & \textbf{44.737} & -16.618 & -11.671 \\
            AF-Low & -186.106 & 12.424 & 43.670 & -40.795 & -201.405 \\
            AF-High & 23.651 & 11.262 & 43.520 & -17.958 & -13.173 \\
            Min-Latency & 28.379 & 11.488 & 43.719 & -16.643 & -10.186 \\
        \midrule
            2RS & 36.757 & 13.728 & 44.967 & -15.334 & -6.603 \\
            GA & 35.914 & 13.253 & 44.797 & -15.442 & -6.683 \\
         \bottomrule
         \multicolumn{6}{l}{* The best method among \textit{distributed optimization methods} is marked \textbf{bold}.}
    \end{tabular}
    }
\end{table}

\paragraph*{High Fairness Scenario}

Here, we analyze the HAF performance for the high fairness scenario, where the distribution of the user's $\alpha$ value is more concentrated in the high regime. 
In this experiment set, we assume each user's $\alpha$ is drawn from the ratio of $\mathcal{A}_1:\mathcal{A}_2:\mathcal{A}_3:\mathcal{A}_4=0.125:0.125:0.375:0.375$.
Similar to the low fairness scenario, we show the HAF and grou-wise HAF in \cref{tab:HAF_table_high_scenario}, and the proposed method outperforms the baseline schemes.
Compared to \cref{tab:HAF_table_normal_scenario}, the HAF of Group $\mathcal{A}_1$ is degraded because the BSs need to allocate more frequency resources to the users with high $\alpha$.

\begin{figure}
    \centering
    \includegraphics[width=0.99\linewidth]{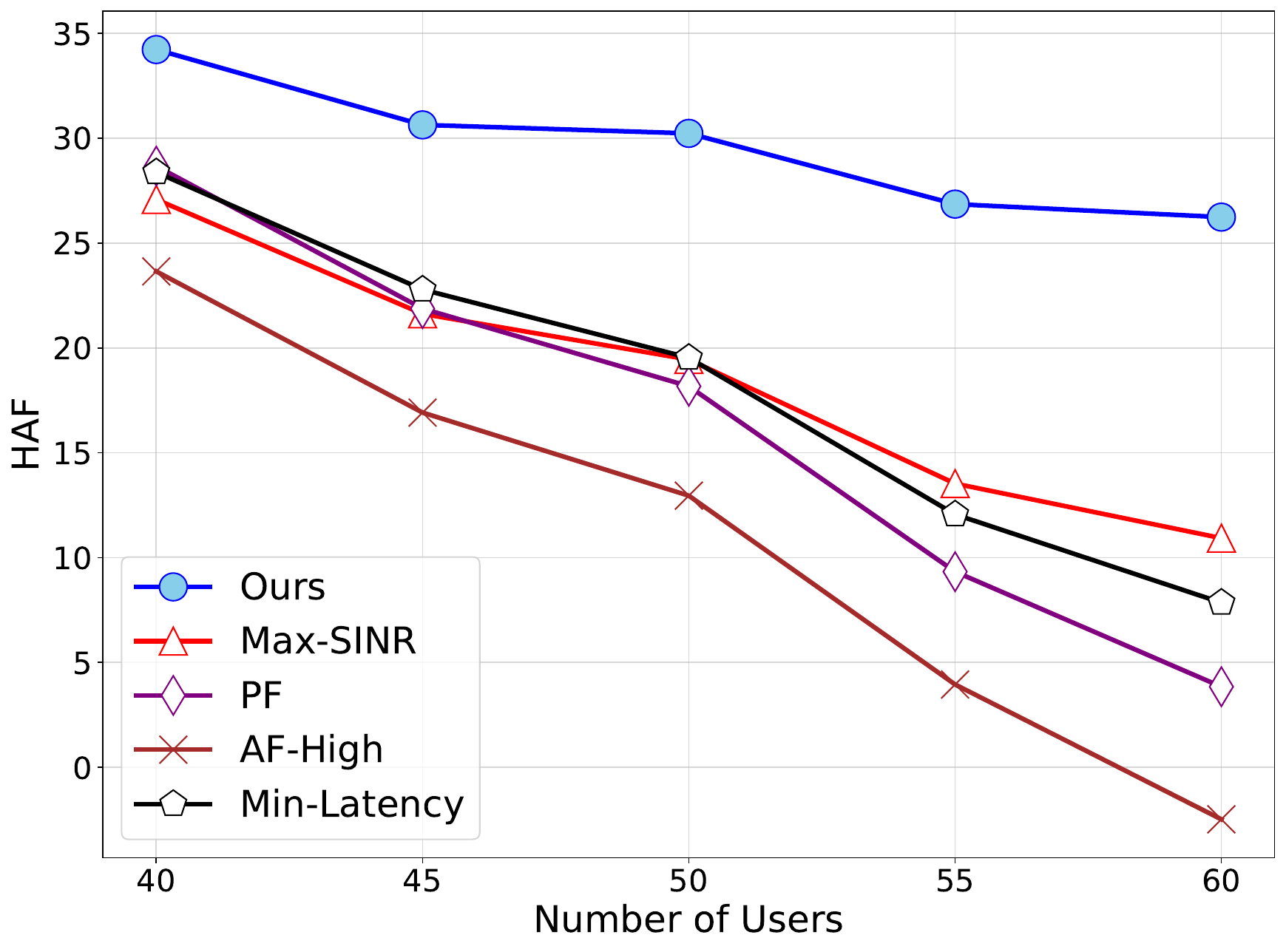}
    \caption{HAF performances of the proposed method and baseline schemes for various numbers of users. }
    \label{fig:HAF_vs_users_high}
\end{figure}

Figure \ref{fig:HAF_vs_users_high} shows the HAF of the proposed method and baselines by varying number of users.
Unlike the result in \cref{fig:HAF_vs_users_normal}, all the methods' HAF decreases as the number of users increases. 
This is because there are more users with $\alpha_i>1$ in the high fairness scenario. As shown in group-wise HAF analysis (\cref{tab:HAF_table_high_scenario}), the HAF of the users $\alpha_i\in[0,1]$ is a positive value, whereas it is negative value if $\alpha_i\in(1,\infty)$. 
Thus, in this scenario, most of the users have $\alpha_i>1$; hence, the HAF tends to decrease with more users.
Despite the high $\alpha$ values of users, with more users, the gap between the proposed method and the baselines increases.

\begin{figure}[tb]
    \centering
    \subfloat[Sum-rate ($\mathcal{A}_1$).\label{subfig:metrics_high_40_a}]{\includegraphics[width=\if 1\doublecolumn .49 \else 0.6 \fi\linewidth]{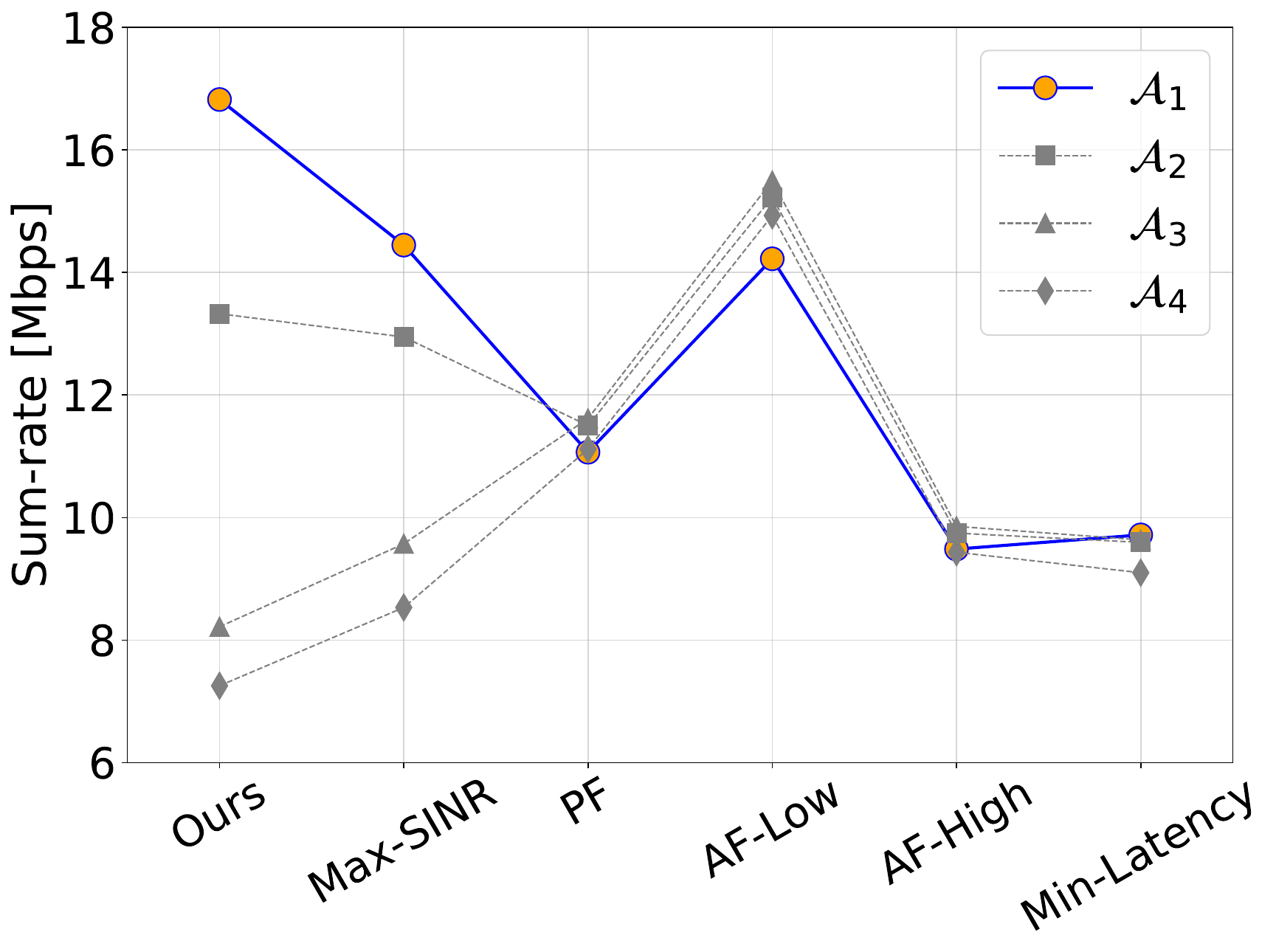}}
    \subfloat[Proportional Fairness ($\mathcal{A}_2$).\label{subfig:metrics_high_40_b}]{\includegraphics[width=\if 1\doublecolumn .49 \else 0.6 \fi\linewidth]{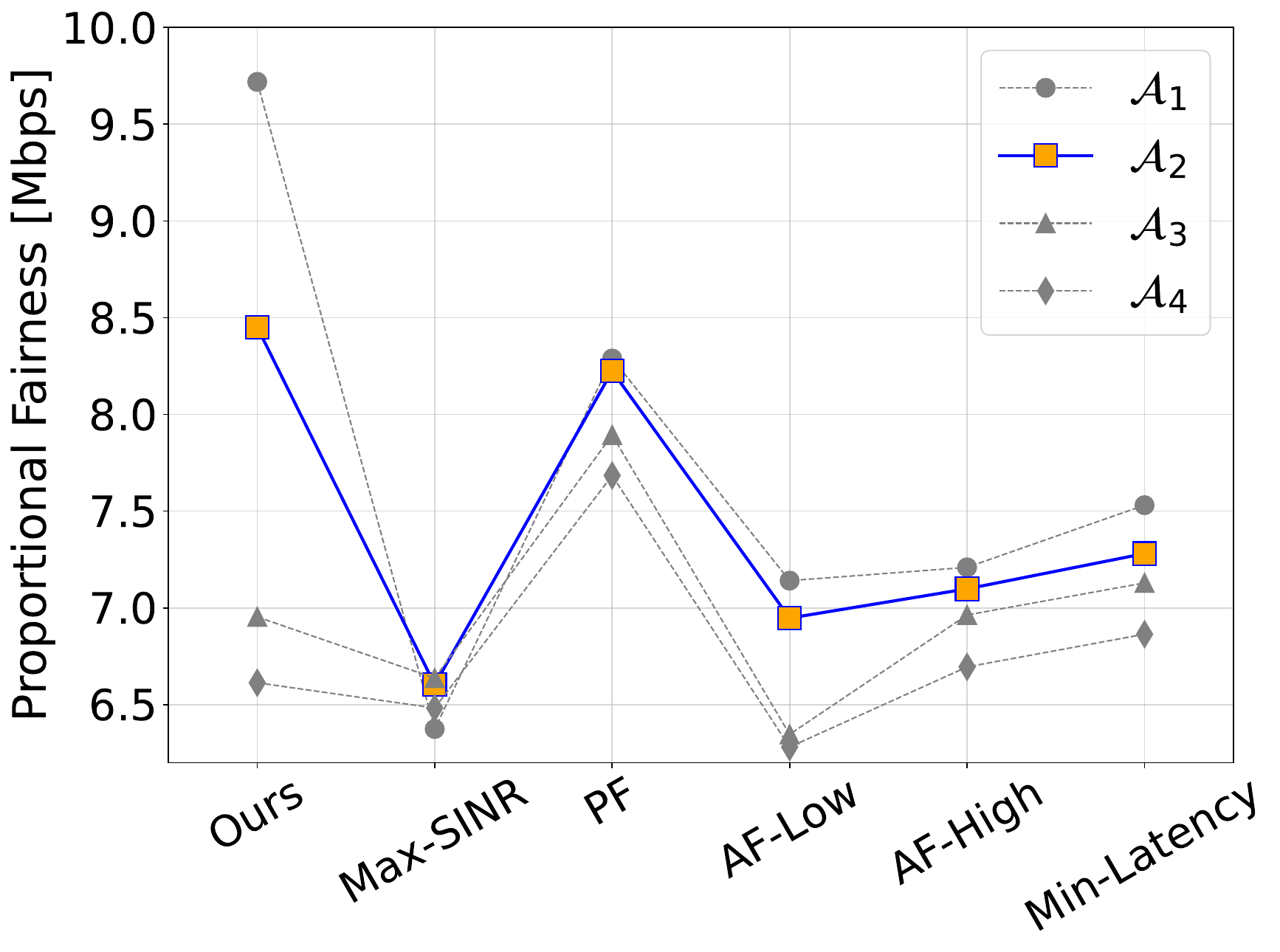}}\\
    \subfloat[Average Latency ($\mathcal{A}_3,\mathcal{A}_4$).\label{subfig:metrics_high_40_c}]{\includegraphics[width=\if 1\doublecolumn .49 \else 0.6 \fi\linewidth]{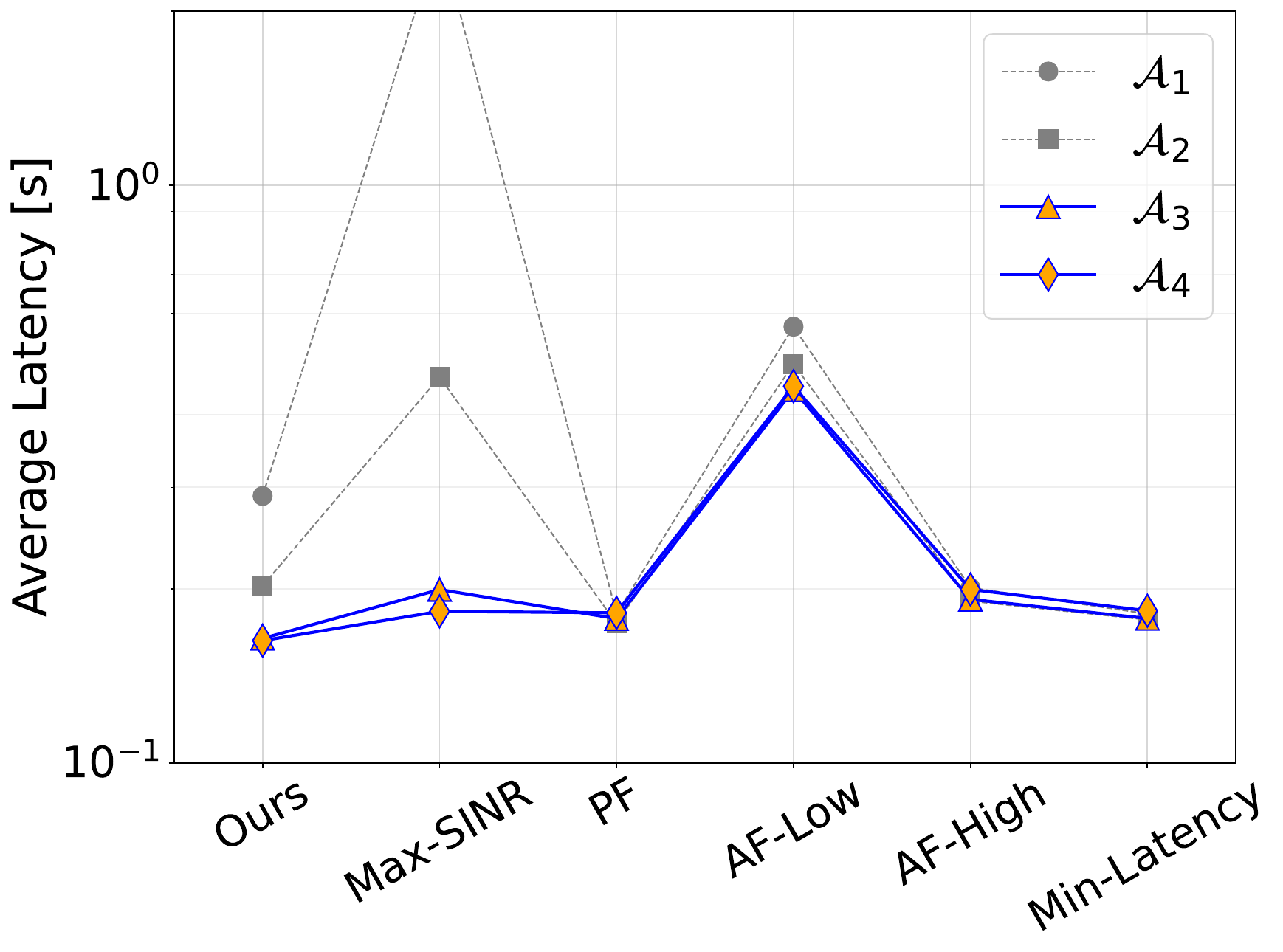}}
    \subfloat[Min-Rate ($\mathcal{A}_4$).\label{subfig:metrics_high_40_d}]{\includegraphics[width=\if 1\doublecolumn .49 \else 0.6 \fi\linewidth]{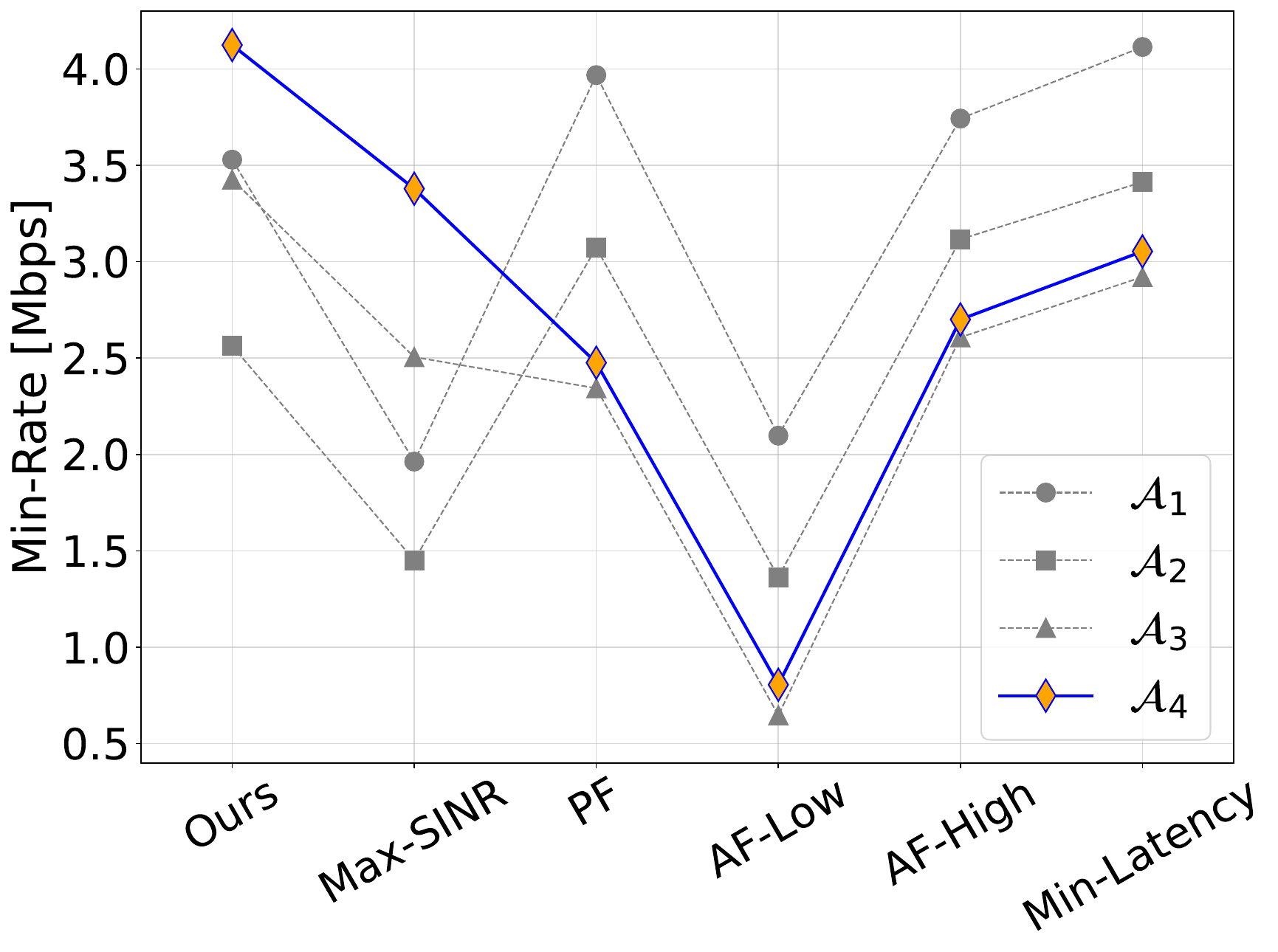}}
    
    \caption{Per-group metrics of the proposed method and baseline schemes: \protect\subref*{subfig:metrics_high_40_a} Sum-rate, \protect\subref*{subfig:metrics_high_40_b} Proportional fairness, \protect\subref*{subfig:metrics_high_40_c} Average latency, and \protect\subref*{subfig:metrics_high_40_d} Min-rate. 
    The groups corresponding to the metric are highlighted in \textbf{blue solid line}, whereas the other groups are represented by \textbf{gray dashed line}.
    }
    \label{fig:metrics_high_40}
\end{figure}

For deeper understanding, we depict the \subref*{subfig:metrics_high_40_a} sum-rate, \subref*{subfig:metrics_high_40_b} proportional fairness, \subref*{subfig:metrics_high_40_c} average latency, and \subref*{subfig:metrics_high_40_d} min-rate of eachgroup in \cref{fig:metrics_high_40}.
Similar to the results in \cref{fig:metrics_normal_40}, the proposed method outperforms the baseline schemes for all baseline methods. 
Unlike \cref{fig:metrics_normal_40}, the optimization of UARA gets more important in the high $\alpha$ scenario, because the optimization with higher $\alpha$ is more sensitive compared to low $\alpha$ scenario (Consider an extreme case $\alpha=0$).
Henceforth, there is a larger room for performance enhancements in the baseline schemes. 
As a result, the proposed method dominates all the baselines for all metrics. 

\subsection{Time-Varying Channel Model}

\begin{figure*}[tb]
    \centering
    \subfloat[Channel correlation of 0.97.\label{subfig:time_slot_003}]{\includegraphics[width=\if 1\doublecolumn .45 \else 0.6 \fi\linewidth]{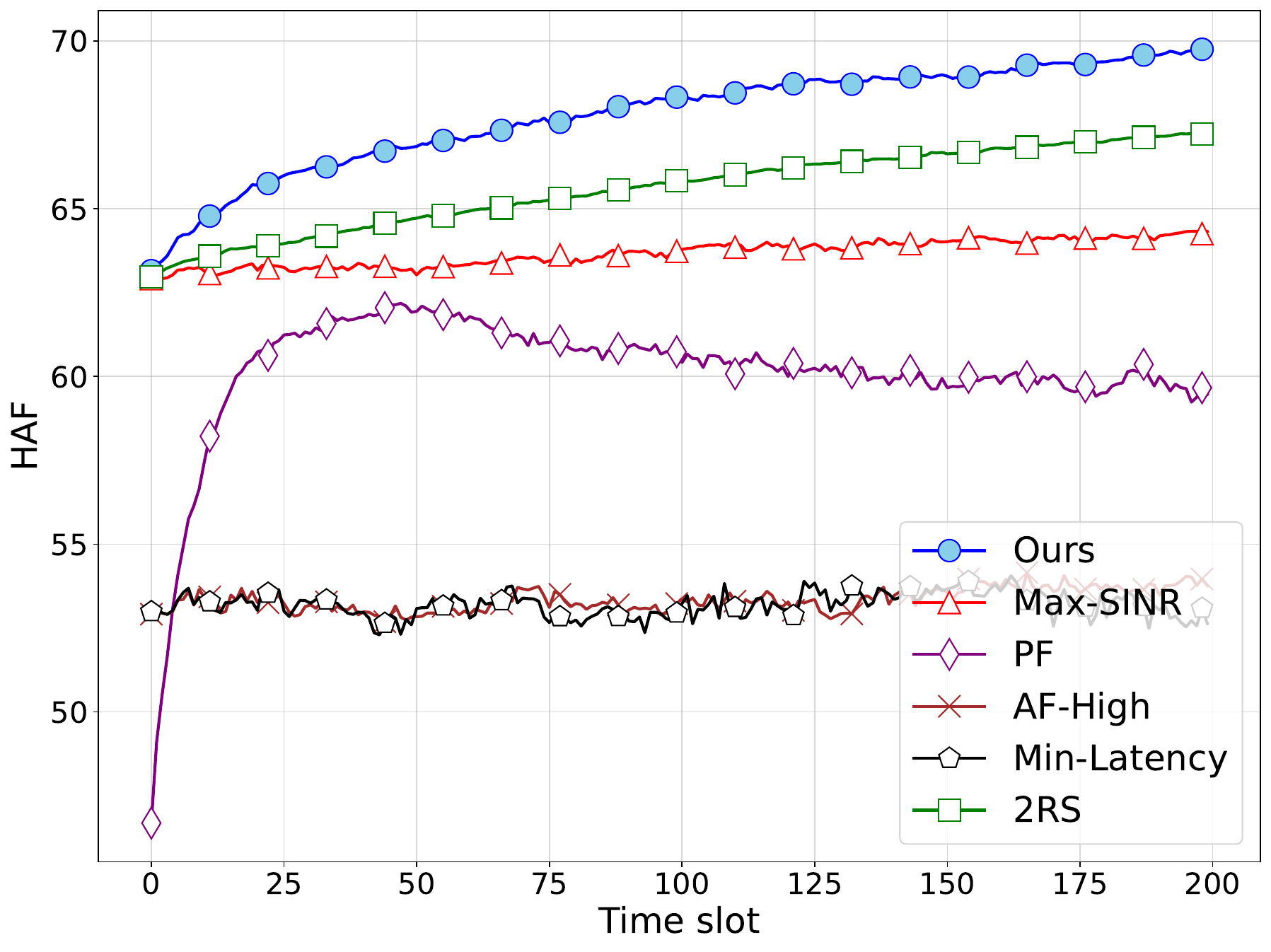}}
    \hfill
    \subfloat[Channel correlation of 0.9.\label{subfig:time_slot_010}]{\includegraphics[width=\if 1\doublecolumn .45 \else 0.6 \fi\linewidth]{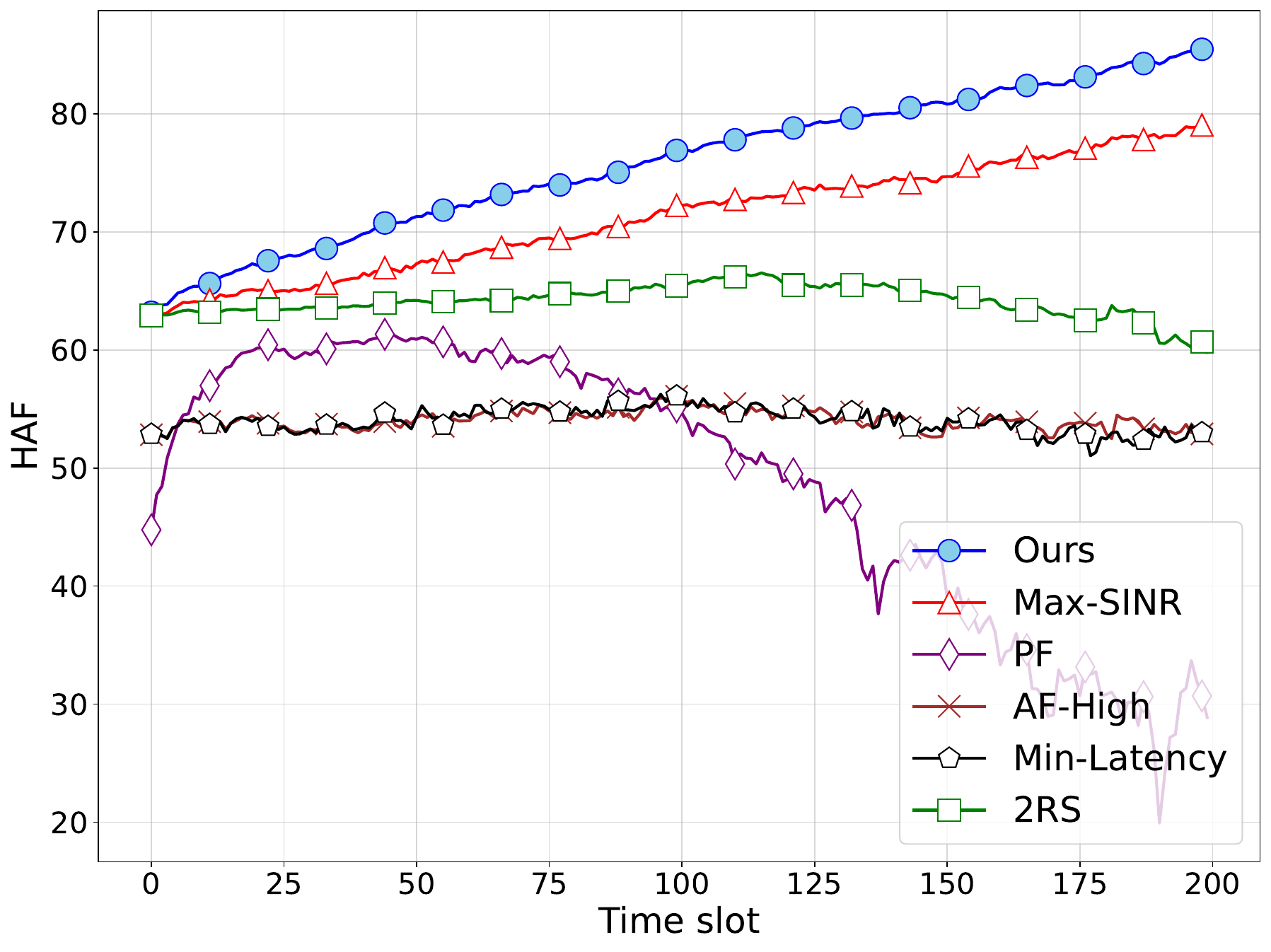}}
    
    \caption{HAF in time-varying channels, where the correlation of the adjacent channel is \protect\subref*{subfig:time_slot_003} 0.97 and \protect\subref*{subfig:time_slot_010} 0.9.
    At the start of the time slot, the price of each method is initialized as a pre-defined constant for fair comparison. 
    In this figure, we consider the low fairness scenario.
    }
    \label{fig:time_slot}
\end{figure*}

\begin{figure*}[tb]
    \centering
    \subfloat[Sum-rate ($\mathcal{A}_1$) with channel correction of 0.97.\label{subfig:metrics_TV_0.03_sum_rate}]{\includegraphics[width=\if 1\doublecolumn .24 \else 0.6 \fi\linewidth]{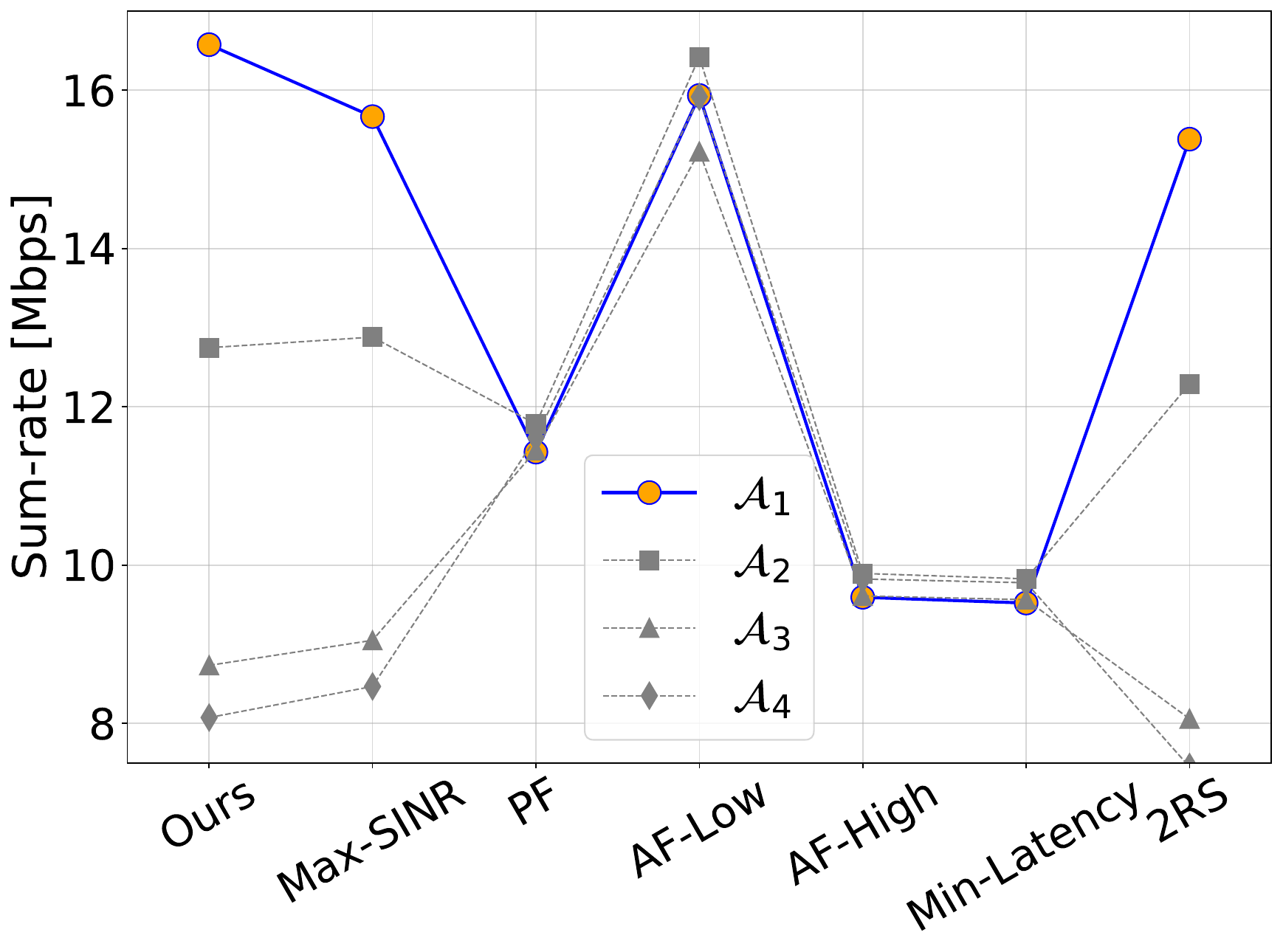}}
    \hfill 
    \subfloat[Proportional fairness ($\mathcal{A}_2$) with channel correction of 0.97.\label{subfig:metrics_TV_0.03_PF}]{\includegraphics[width=\if 1\doublecolumn .24 \else 0.6 \fi\linewidth]{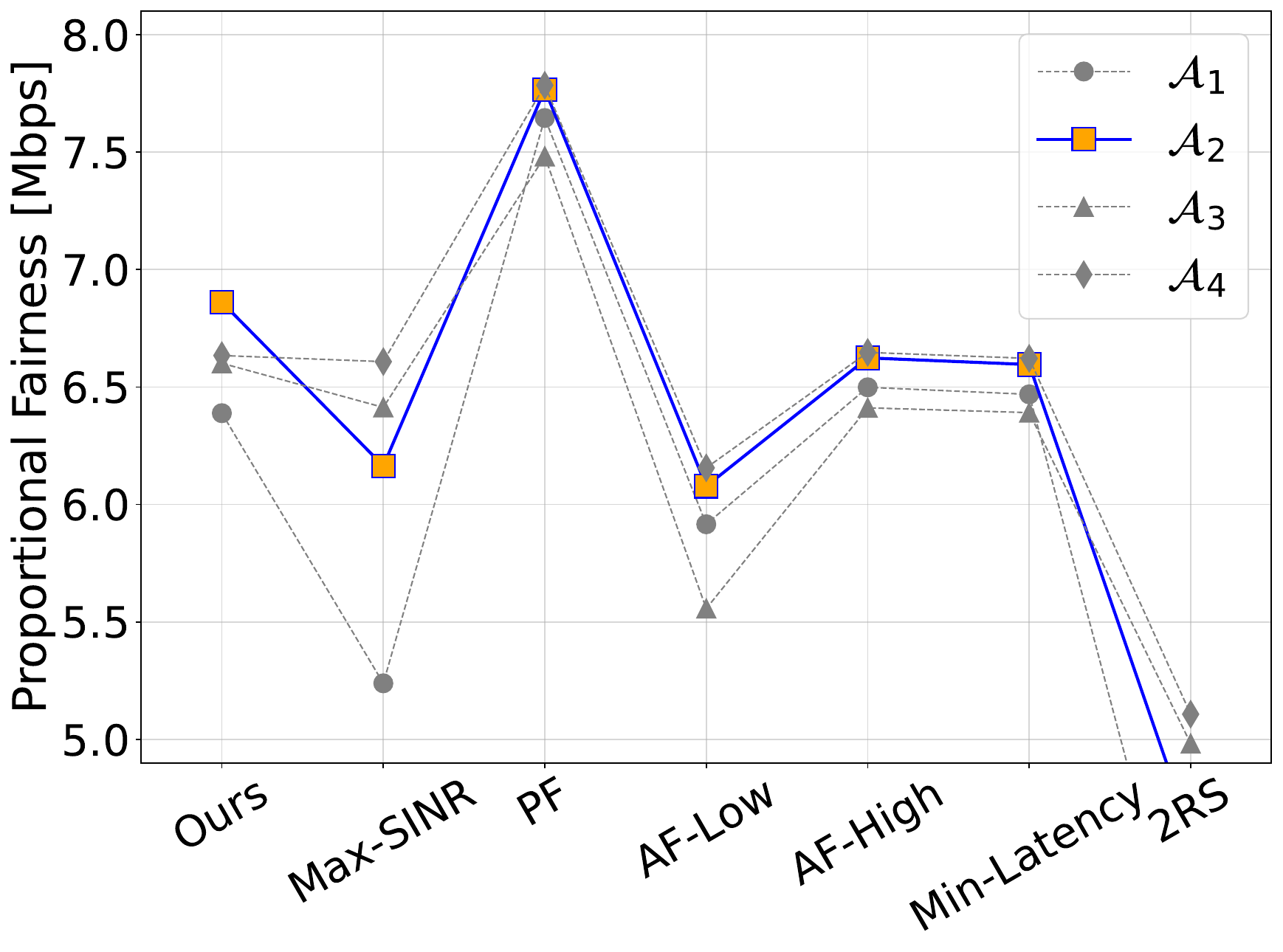}}
    \hfill 
    \subfloat[Average latency ($\mathcal{A}_3,\mathcal{A}_4$) with channel correction of 0.97.\label{subfig:metrics_TV_0.03_latency}]{\includegraphics[width=\if 1\doublecolumn .24 \else 0.6 \fi\linewidth]{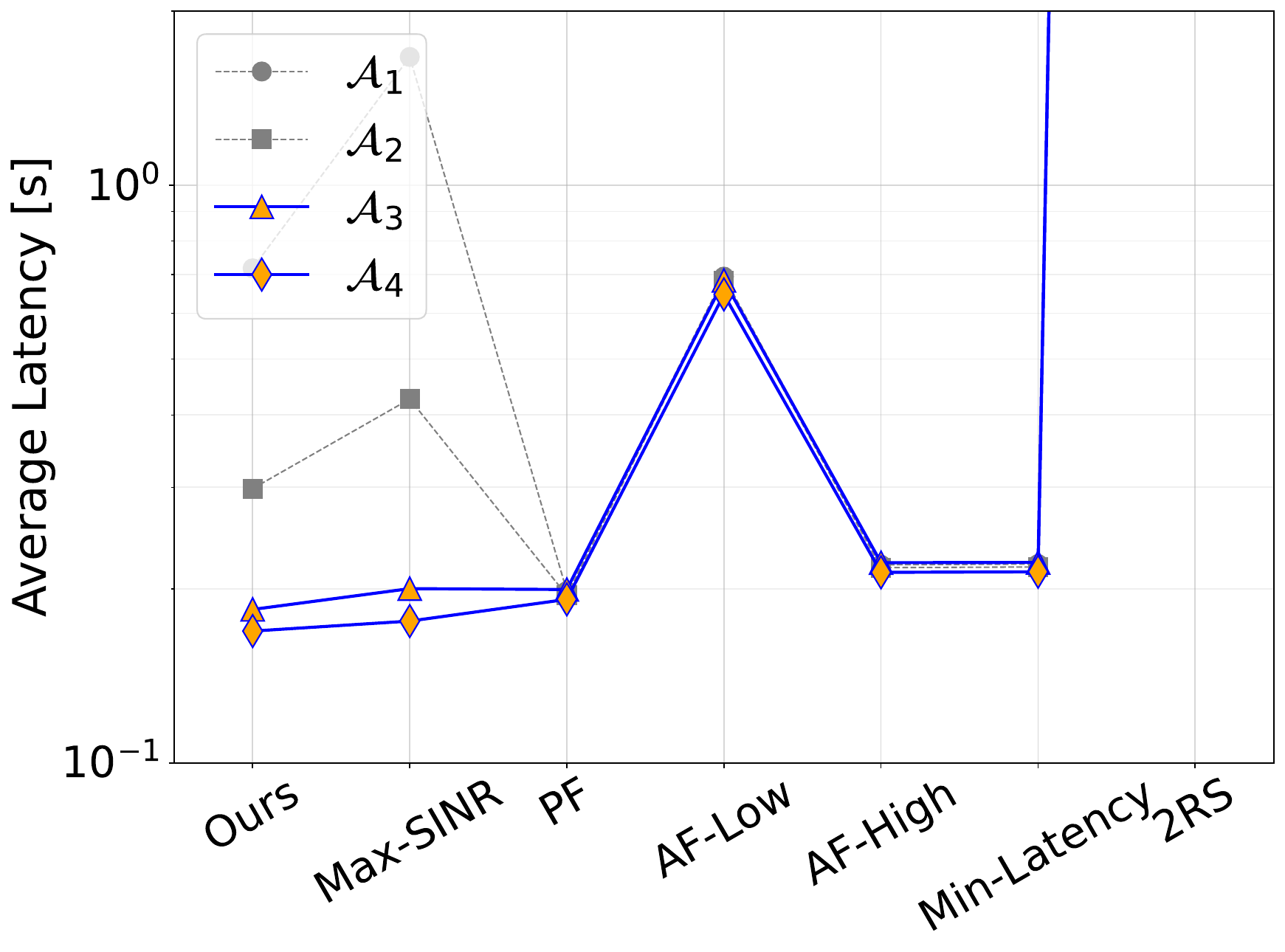}}
    \hfill
    \subfloat[Min-rate ($\mathcal{A}_4$) with channel correction of 0.97.\label{subfig:metrics_TV_0.03_min_rate}]{\includegraphics[width=\if 1\doublecolumn .24 \else 0.6 \fi\linewidth]{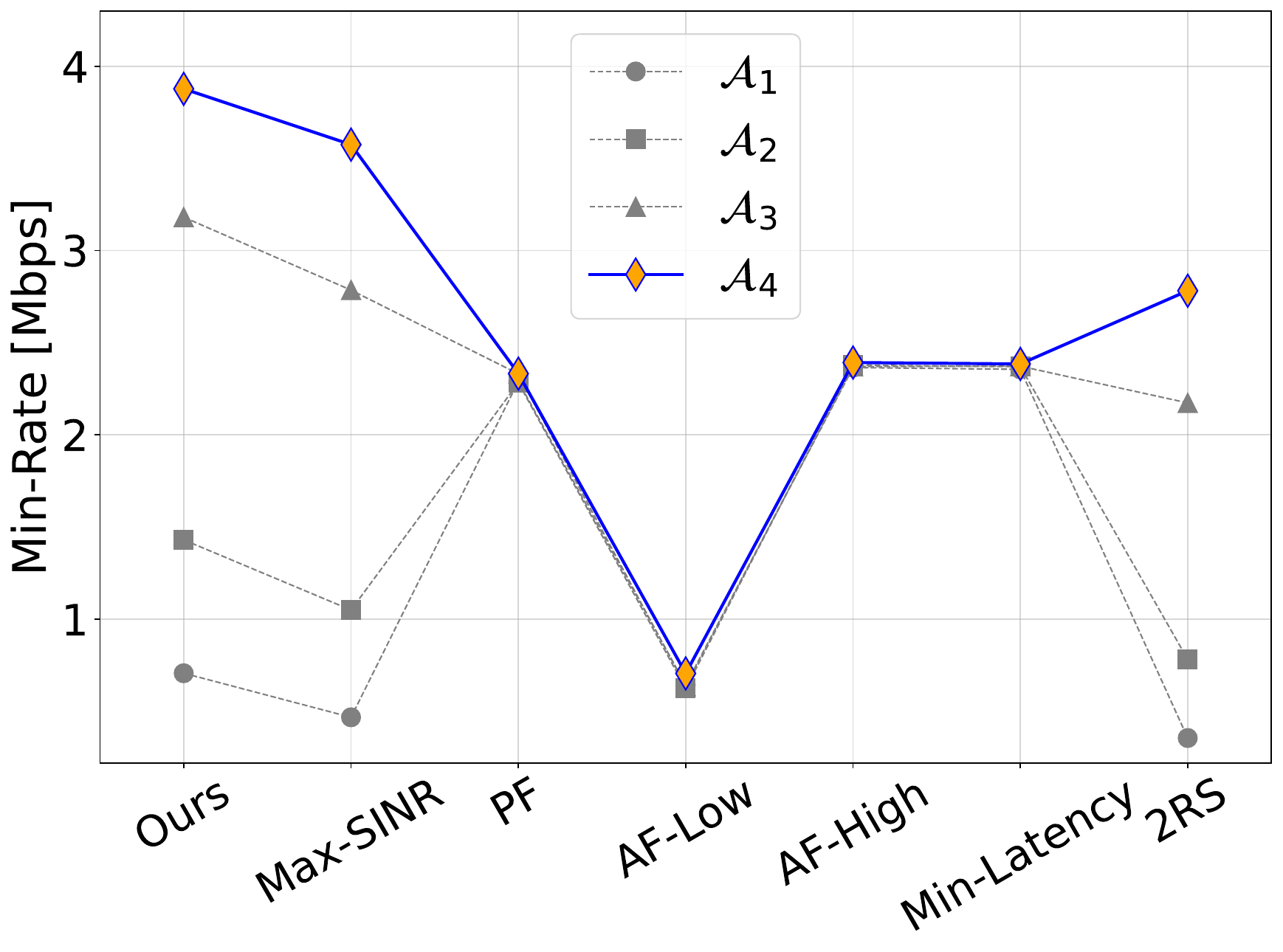}}\\
    \subfloat[Sum-rate ($\mathcal{A}_1$) with channel correction of 0.9.\label{subfig:metrics_TV_0.1_sum_rate}]{\includegraphics[width=\if 1\doublecolumn .24 \else 0.6 \fi\linewidth]{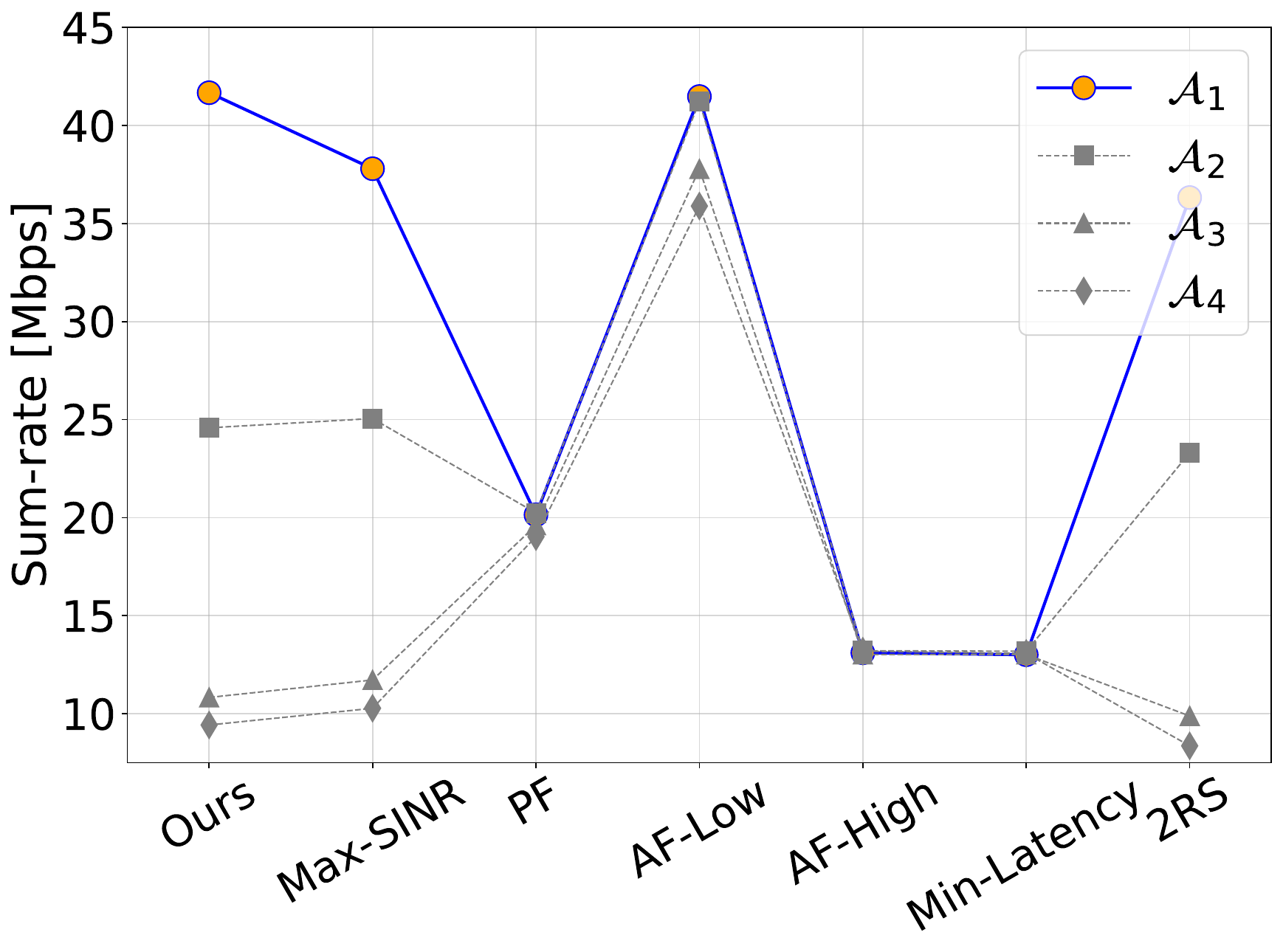}}
    \hfill 
    \subfloat[Proportional fairness ($\mathcal{A}_2$) with channel correction of 0.9.\label{subfig:metrics_TV_0.1_PF}]{\includegraphics[width=\if 1\doublecolumn .24 \else 0.6 \fi\linewidth]{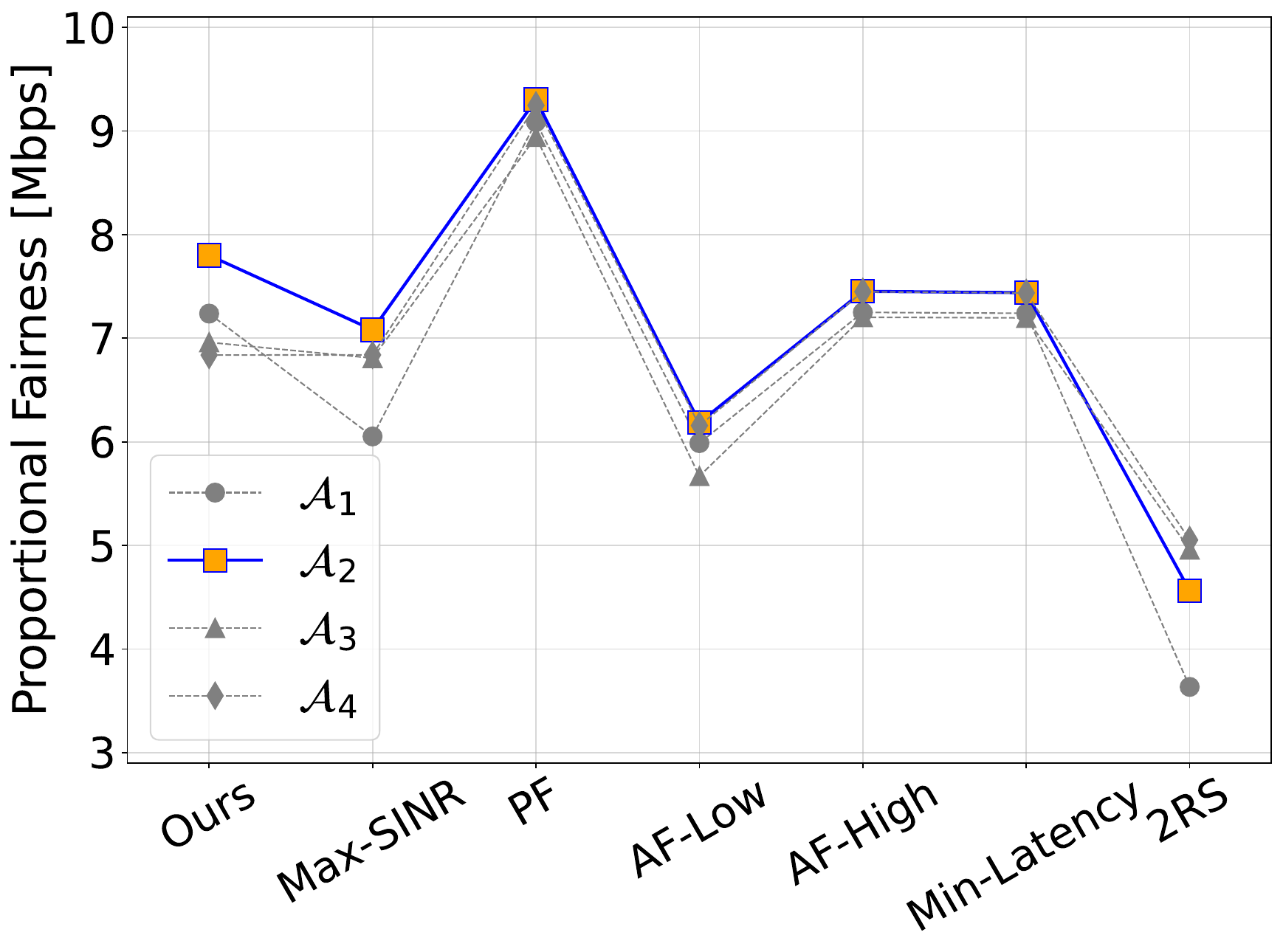}}
    \hfill 
    \subfloat[Average latency ($\mathcal{A}_3,\mathcal{A}_4$) with channel correction of 0.9.\label{subfig:metrics_TV_0.1_latency}]{\includegraphics[width=\if 1\doublecolumn .24 \else 0.6 \fi\linewidth]{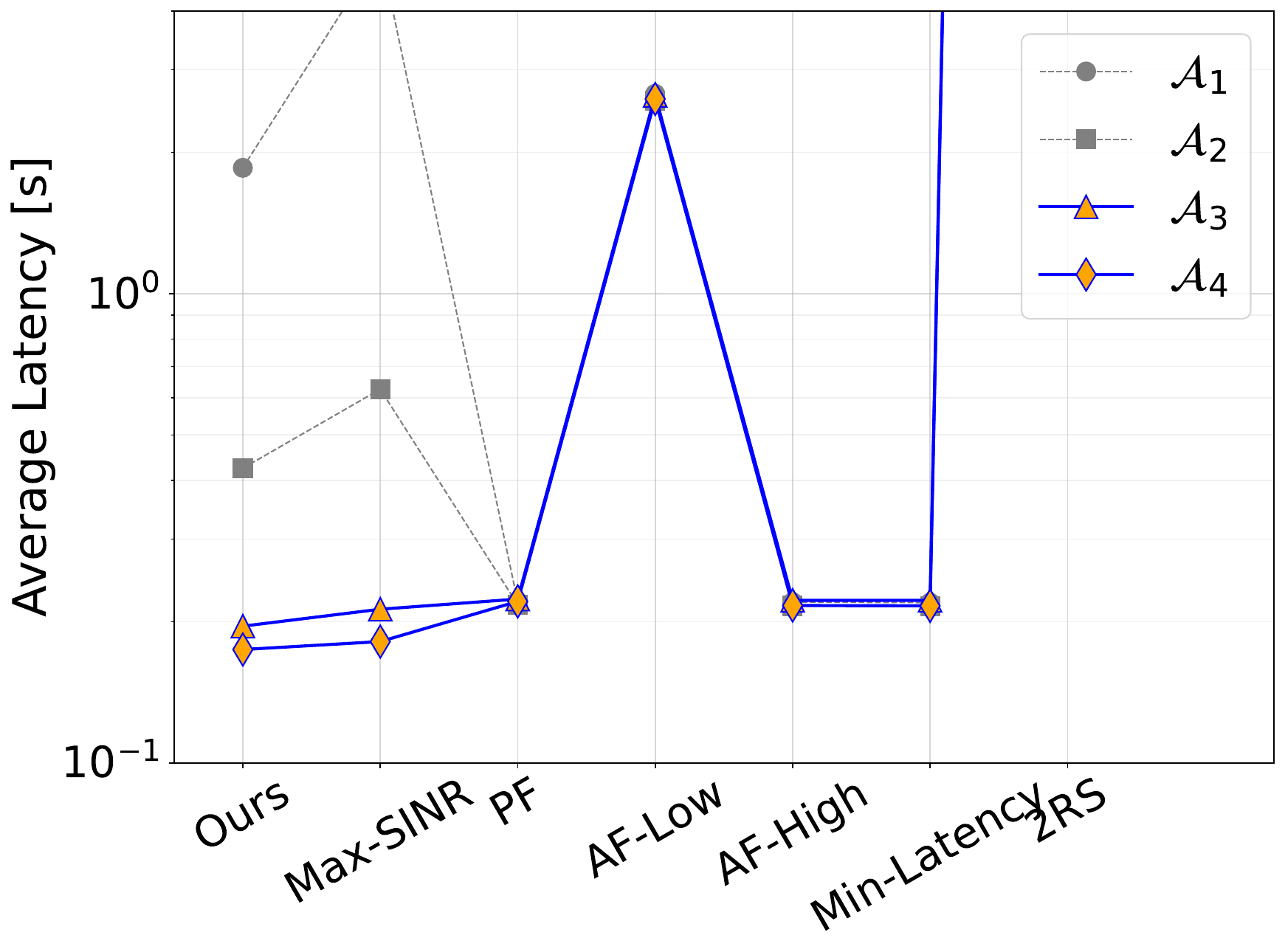}}
    \hfill
    \subfloat[Min-rate ($\mathcal{A}_4$) with channel correction of 0.9.\label{subfig:metrics_TV_0.1_min_rate}]{\includegraphics[width=\if 1\doublecolumn .24 \else 0.6 \fi\linewidth]{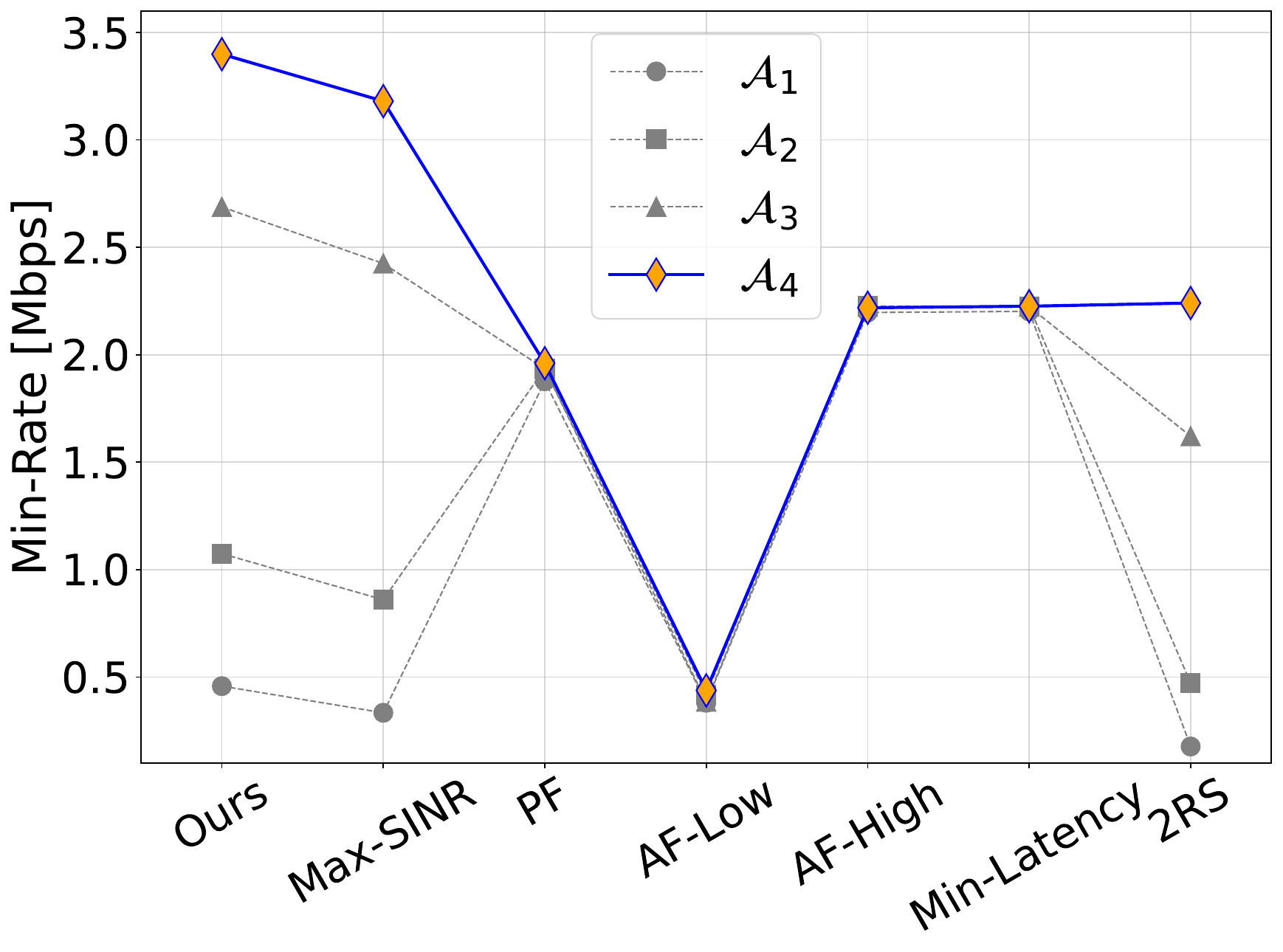}}
    \caption{Per-group metrics of the proposed method and baseline schemes (\protect\subref*{subfig:metrics_TV_0.03_sum_rate},\protect\subref*{subfig:metrics_TV_0.1_sum_rate}: Sum-rate, \protect\subref*{subfig:metrics_TV_0.03_PF},\protect\subref*{subfig:metrics_TV_0.1_PF}: Proportional fairness, \protect\subref*{subfig:metrics_TV_0.03_latency}, \protect\subref*{subfig:metrics_TV_0.1_latency}: Average latency, and \protect\subref*{subfig:metrics_TV_0.03_min_rate}, \protect\subref*{subfig:metrics_TV_0.1_min_rate}: Min-rate). 
    In this figure, we consider two time-varying channel models (\protect\subref*{subfig:metrics_TV_0.03_sum_rate}-\protect\subref*{subfig:metrics_TV_0.03_min_rate}: channel correlation of 0.97 and \protect\subref*{subfig:metrics_TV_0.1_sum_rate}-\protect\subref*{subfig:metrics_TV_0.1_min_rate}: channel correlation of 0.9).
    The groups corresponding to the metric are highlighted in \textbf{a blue solid line}, whereas the other groups are represented by \textbf{a gray dashed line}.
    }
    \label{fig:metric_TV}
\end{figure*}

In this section, we evaluate the proposed method in time-varying channels. 
Because the proposed method is a pricing-based optimization approach, it is a kind of adaptive method; hence, we need to discuss the impact of time-varying channels. 
For comparison, we add a modified 2-distance ring solution (2RS), where it only takes a single iteration in each time slot for fairness. 
In \cref{fig:time_slot}, we depict the HAF of the proposed methods and baseline schemes in time-varying channels, where the correlation of the adjacent channel is 0.97 in \cref{subfig:time_slot_003} and 0.9 in \cref{subfig:time_slot_010}. 
In the overview of the figure, the proposed method still outperforms the baseline schemes in both time-varying channel scenarios. 
More importantly, by comparing \cref{subfig:time_slot_003} and \cref{subfig:time_slot_010}, the proposed method still adapts the channel condition; however, the 2RS scheme, which can only change one user's association, fails on the optimization in \cref{subfig:time_slot_010} because of the highly varying channels.

To further evaluate the proposed method with various metrics, we depict \protect\subref*{subfig:metrics_TV_0.03_sum_rate},\protect\subref*{subfig:metrics_TV_0.1_sum_rate}: sum-rate, \protect\subref*{subfig:metrics_TV_0.03_PF},\protect\subref*{subfig:metrics_TV_0.1_PF}: proportional fairness, \protect\subref*{subfig:metrics_TV_0.03_latency}, \protect\subref*{subfig:metrics_TV_0.1_latency}: average latency, and \protect\subref*{subfig:metrics_TV_0.03_min_rate}, \protect\subref*{subfig:metrics_TV_0.1_min_rate}: min-rate in \cref{fig:metric_TV}. 
Similar to the previous results in \cref{fig:metrics_normal_40,fig:metrics_high_40}, the proposed method consistently outperforms the baseline pricing-based methods as well as Max-SINR and 2RS schemes. 

By doing the above experiments, we confirm that the proposed method well adapts even to the time-varying channels. whereas the HAF of the 2RS baseline scheme is degraded as the channel correlation decreases.

\subsection{Computational Complexity}

Let us denote the computational complexity of the RA optimization as $K$, where the computational complexity of Algorithm \ref{algo:ra_algorithm} is $\mathtt{iters}\cdot3\cdot\mathcal{O}(JI)$.
Then, the pricing-based schemes (Ours, PF, AF-Low, AF-High) require $K + \mathcal{O}(JI)$ flops per iteration. 
On the other hand, the 2RS scheme requires $K\cdot\mathcal{O}(JI)$ flops per iteration, where the adaptive mode of this scheme requires $K$ flops per iteration. 
Another centralized optimization scheme, GA, requires $G\cdot P \cdot K$ iterations, where $G$ and $P$ denote the number of generations and populations, respectively.

\section{Discussion and Conclusion}



In summary, we have proposed a novel heterogeneous alpha-fairness (HAF) framework for joint user association and resource allocation. Our distributed pricing-based algorithm achieves flexible prioritization across users, adapts to varying channel conditions, and outperforms conventional fairness models.
Also, we showed the theoretical convergence and optimality analysis of the proposed method. 
Our results demonstrated that the proposed method nearly achieves the upper bound obtained by the theoretical analysis (in \cref{fig:convergence}) and outperforms the baseline schemes in various scenarios. 
Our analysis and experiments indicate that optimizing HAF can enhance the network-wide performance by allocating frequency resources (RA) to appropriate users (UA), shown in \cref{fig:metrics_normal_40,fig:metrics_high_40,fig:metric_TV}. 
Also, because the proposed method is based on the pricing-based optimization, it is easy to implement, as it does not require complex implementations, thereby raising interest in practical usages. 
Despite these advantages, the current formulation assumes static user demands and single-antenna systems. Future work includes integrating MIMO schemes and dynamic QoS-aware fairness adaptation.

\appendices

\section{KKT Condition Analysis of RA Optimiation\label{sec:appendix_KKT_RA}}

The Lagrangian of Problem \ref{eq:P3} is represented by 
\begin{equation}
\begin{split}
    L_{\mathcal{P}\text{3}} = & \sum_{i\in\mathcal{I}_j} \frac{(\gamma_{ij}y_{ij})^{1-\alpha_i}}{1-\alpha_i} + \lambda_j\left(1 - \sum_{i\in\mathcal{I}_j} y_{ij}\right) \\ & + \sum_{i\in\mathcal{I}_j}\xi_i y_{ij},
\end{split}
\end{equation}
where $\lambda$ and $\xi_i$ denote the Lagrangian multipliers w.r.t. the constraints \eqref{eq:P3_b} and \eqref{eq:P3_c}, respectively.
Then, the KKT conditions of Problem \ref{eq:P3} is derived as 
\begin{equation}
    \begin{cases}
        \gamma_{ij}^{1-\alpha_i}y_{ij}^{-\alpha_i} = \lambda_j - \xi_i , & \forall i\in\mathcal{I}_j  \\ 
        \sum_{i\in\mathcal{I}_j} y_{ij} \le 1 \\ 
        y_{ij} \le 0 \\ 
        \lambda_j \left(1 - \sum_{i\in\mathcal{I}_j} y_{ij} \right) = 0 \\ 
        \xi_j y_{ij}= 0, & \forall i\in\mathcal{I}_j \\ 
        \lambda_j \ge 0 \\ 
        \xi_i \ge 0, & \forall i\in\mathcal{I}_j.
    \end{cases}
\end{equation}

Hereafter, our focus is to find an optimal solution that satisfies the conditions. 
We first divide the cases of the condition by i) $\xi_i>0$ for some $i\in\mathcal{I}_j$ and ii) $\xi_i=0$ for all $i\in\mathcal{I}_j$.

\paragraph*{Case 1}
If $\xi_i>0$ for some $i\in\mathcal{I}_j$, it means there exists an index $i$ that satisfies $y_{ij}=0$. 
However, from the first KKT condition, there exists no $\lambda_j$ satisfying $(\lambda_j -\xi_i)y_{ij}^{\alpha_i} = \gamma_{ij}^{1-\alpha_i}$, because $(\lambda_j - \xi_i)y_{ij}^{\alpha_i}=0$ if $\alpha_i>0$ and $\gamma_{ij}^{1-\alpha_i}>0$. 
That is, this case is infeasible. 

\paragraph*{Case 2}
Because the first case does not provide a feasible solution, we consider the case where $\xi_i=0$ for all $i\in\mathcal{I}_j$. 
If $\xi_i=0$, from the first condition, we have
\begin{equation}
    y_{ij} = \gamma_{ij}^{\frac{1}{\alpha_i} - 1 } \lambda_j^{-\frac{1}{\alpha_i}},
\end{equation}
where the value of $\lambda_j\neq0$ to have a feasible solution because $\alpha_i>0$. 
If $\lambda_j>0$, all the KKT conditions except the fourth condition are satisfied.
By considering the fourth condition, we need to find a solution $y_{ij}$ satisfying $\sum_{i\in\mathcal{I}_j}y_{ij} = 1$. 

Thus, the KKT condition of Problem \ref{eq:P3} implies that finding $\lambda_j$ satisfies the following condition is equivalent to finding the optimal solution of Problem \ref{eq:P3}:
\begin{equation}\label{eq:appendix_lambda}
       \sum_{i\in\mathcal{I}_j} \lambda_j^{-\frac{1}{\alpha_i}}\gamma_{ij}^{\frac{1}{\alpha_i}-1} = 1.
\end{equation}
Because $i\in\mathcal{I}_j$ if $x_{ij} = 1$, and since $x_{ij}\in\{0,1\}$, the condition \eqref{eq:appendix_lambda} can be rewritten by 
\begin{equation}\label{eq:appendix_lambda_2}
   \sum_{i\in\mathcal{I}} \lambda_j^{-\frac{1}{\alpha_i}} \gamma_{ij}^{\frac{1}{\alpha_i}-1}x_{ij} = 1. 
\end{equation}


\section{Proof of Theorem \ref{thm:convergence}}
\label{appendix:convergence}

In this appendix, we prove the convergence of Algorithm \ref{algo:ua_algorithm}. 
For the proof, we denote the optimal solution of Problem \ref{eq:P5} as $\boldsymbol{\mu}^*$. 
Also, we assume $\Vert\mathbf{g}_t\Vert \le G$ for all $t\in\mathbb{N}$, where $[\mathbf{g}]_i = 1 - \widehat{\gamma}_{ij}\mu_j^{-\frac{1}{\alpha_i}}$
Because the objective function $g(\boldsymbol{\mu})$ is convex w.r.t. $\boldsymbol{\mu}$, we have
\begin{equation}\label{eq:appendix_concave}
    \begin{split}
        g(\boldsymbol{\mu}_1) - g(\boldsymbol{\mu}_2) \le \mathbf{g}_1^\mathrm{T}(\boldsymbol{\mu}_1 - \boldsymbol{\mu}_2), ~ \forall \mathbf{g}_2\in\partial g(\boldsymbol{\mu}_2),
    \end{split}
\end{equation}
where $\partial g(\boldsymbol{\mu}^{(t)})$ denotes a set of subgradients of $g(\cdot)$ at $\boldsymbol{\mu}^{(t)}$.
In \eqref{eq:subgrad}, the price $\boldsymbol{\mu}$ is updated by 
\begin{equation}
    \boldsymbol{\mu}^{(t+1)}  = [\boldsymbol{\mu}^{(t)} - \eta \mathbf{g}_t^\mathrm{T}]_+,
\end{equation}
where $[\cdot]_+$ denotes $\max(0, \cdot)$.
Then, because the optimal solution $\boldsymbol{\mu}\ge \mathbf{0}$, we have
\begin{equation}\label{eq:appendix_ineq_thm1}
    \begin{split}
        \Vert\boldsymbol{\mu}^{(t+1)} - \boldsymbol{\mu}^*\Vert^2 & = \Vert [\boldsymbol{\mu}^{(t)} - \eta \mathbf{g}_t^\mathrm{T}]_+ - \boldsymbol{\mu}^* \Vert^2 \\
        & \le \Vert \boldsymbol{\mu}^{(t)} - \eta \mathbf{g}_t^\mathrm{T} - \boldsymbol{\mu}^* \Vert^2 \\ 
        & =  \Vert \boldsymbol{\mu}^{(t)} - \boldsymbol{\mu}^* \Vert^2 + \eta^2 \Vert \mathbf{g}_t\Vert^2 \\ 
        & ~~~~~ - 2\eta \mathbf{g}_t^\mathrm{T}\left(\boldsymbol{\mu}^{(t)} - \boldsymbol{\mu}^*\right).
    \end{split}
\end{equation}
By substituting \eqref{eq:appendix_ineq_thm1} into \eqref{eq:appendix_concave}, we have
\begin{equation}
    \begin{split}
        2\eta\left( g(\boldsymbol{\mu}^{(t)}) - g(\boldsymbol{\mu}^*)\right) & \le 
        2\eta \mathbf{g}_t^\mathrm{T}\left(\boldsymbol{\mu}^{(t)} - \boldsymbol{\mu}^*\right) \\ 
        & \le \Vert\boldsymbol{\mu}^{(t)}-\boldsymbol{\mu}^*\Vert^2 + \eta^2\Vert\mathbf{g}_t\Vert^2 \\ 
        & ~~~~~ - \Vert\boldsymbol{\mu}^{(t+1)}-\boldsymbol{\mu}^*\Vert^2,
    \end{split}
\end{equation}
where the second inequality is obtained from \eqref{eq:appendix_ineq_thm1}.
Because our focus is to derive the convergence of $\min_{t\in T} g(\boldsymbol{\mu}^{(t)}) - g(\boldsymbol{\mu}^*)$, we represent the convergence of Algorithm \ref{algo:ua_algorithm} as
\begin{equation}
    \begin{split}
        \min_{t\in T} g(\boldsymbol{\mu}^{(t)}) - g(\boldsymbol{\mu}^*) & \le \frac{1}{T}\sum_{t=1}^{T} \left(g(\boldsymbol{\mu}^{(t)}) - g(\boldsymbol{\mu}^*) \right)\\ 
        & \le \frac{\Vert\boldsymbol{\mu}^{(1)} - \boldsymbol{\mu}^* \Vert^2}{2T\eta} + \frac{\eta}{2T}\sum_{t=1}^{T}\Vert\mathbf{g}_t\Vert^2\\ 
        & ~~~~~ - \frac{\Vert\boldsymbol{\mu}^{(T+1)} - \boldsymbol{\mu}^* \Vert^2}{2T\eta} \\ 
        & \le \frac{\Vert\boldsymbol{\mu}^{(1)} - \boldsymbol{\mu}^* \Vert^2}{2T\eta} + \frac{\eta}{2}G^2 \\ 
        & \le \frac{G\Vert\boldsymbol{\mu}^{(1)} - \boldsymbol{\mu}^* \Vert^2}{\sqrt{T}},
    \end{split}
\end{equation}
where the last inequality holds if $\eta = \frac{\Vert\boldsymbol{\mu}^{(1)} - \boldsymbol{\mu}^* \Vert}{G\sqrt{T}}$.
Thus, we complete the proof for the convergence of Algorithm \ref{algo:ua_algorithm}. 

\section{Proof of Theorem \ref{thm:optimality}}
\label{sec:appendix_optimality}
Let us denote the solution of Problem \ref{eq:P5} as $\boldsymbol{\mu}^*$. 
Then, from the condition \eqref{eq:optimal_lambda}, we have $\Lambda^*=\boldsymbol{\mu}^*$.
However, it does not strictly indicate that $\Lambda^*$ does not equal the $\Lambda$ obtained from Algorithm \ref{algo:ra_algorithm}. 
Hence, let us define $\Lambda$ obtained from Algorithm \ref{algo:ra_algorithm} as $\widehat{\Lambda}$. 
Our focus is to derive the optimality of the solution obtained by Algorithm \ref{algo:ua_algorithm}.
For brevity of the notation, we let $f^*$ be the HAF obtained by Algorithm \ref{algo:ua_algorithm} as follows:
\begin{equation}
    f^* = \sum_{j\in\mathcal{J}}\sum_{i\in\mathcal{I}}\frac{\hat{\gamma}_{ij}}{1-\alpha_i}\widehat{\lambda}_i^\frac{\alpha_i-1}{\alpha_i}.
\end{equation}

Then, for the brevity of the notation, we let the optimal value of the dual function by $g(\boldsymbol{\mu})$ as $d^*$.
Then, the following inequality holds
\begin{equation}
    \begin{split}\sum_{j\in\mathcal{J}}\sum_{i\in\mathcal{I}}\frac{1}{{1-\alpha_i}} \widehat{\gamma}_{ij}\lambda_i^{\frac{\alpha_i-1}{\alpha_i}}x_{ij} \le d^*,
    \end{split}
\end{equation}
if $\Lambda$ and $\mathbf{X}$ meet
\begin{equation}
    \begin{split}
        \begin{cases}
            \sum_{j\in\mathcal{J}}x_{ij} = 1 \\ 
            x_{ij}\in\{0,1\}, & \forall i\in\mathcal{I},j\in\mathcal{J}\\
            \sum_{k\in\mathcal{I}}\widehat{\gamma}_{kj}\lambda_k^{-\frac{1}{\alpha_k}}x_{kj} =1, & \forall j\in\mathcal{J}.
        \end{cases}
    \end{split}
\end{equation}
Thus, from the weak duality condition, the optimal value of the dual function is an upper bound of the HAF, \ie, $f^* \le g^*$.
Then, we can obtain the optimality gap of Algorithm \ref{algo:ua_algorithm} for the HAF objective function as
\begin{equation}
    \begin{split}
        g^* - f^* & \le \sum_{j\in\mathcal{J}}\lambda_j^* + \sum_{i\in\mathcal{I}}\sum_{j\in\mathcal{J}}\frac{\alpha_i}{1-\alpha_i} \hat{\gamma}_{ij}x_{ij}(\lambda_j^*)^\frac{\alpha_i-1}{\alpha_i} \\ 
        & ~~~~~ - \sum_{j\in\mathcal{J}}\sum_{i\in\mathcal{I}}\frac{\hat{\gamma}_{ij}x_{ij}}{1-\alpha_i}\widehat{\lambda}_i^\frac{\alpha_i-1}{\alpha_i} \\ 
        & =  \sum_{j\in\mathcal{J}}\lambda_j^* + \sum_{i\in\mathcal{I}}\sum_{j\in\mathcal{J}}\frac{\alpha_i}{1-\alpha_i} \hat{\gamma}_{ij}x_{ij}(\lambda_j^*)^\frac{\alpha_i-1}{\alpha_i} \\ 
        & ~~~~~ - \sum_{j\in\mathcal{J}}\widehat{\lambda}_j\sum_{i\in\mathcal{I}}\left(1 + \frac{\alpha_i}{1-\alpha_i}\right)\hat{\gamma}_{ij}x_{ij}\widehat{\lambda}_j^\frac{-1}{\alpha_i} \\ 
        & = \sum_{j\in\mathcal{J}}\lambda_j^* + \sum_{i\in\mathcal{I}}\sum_{j\in\mathcal{J}}\frac{\alpha_i}{1-\alpha_i} \hat{\gamma}_{ij}x_{ij}(\lambda_j^*)^\frac{\alpha_i-1}{\alpha_i} \\ 
        & ~~~~~ - \sum_{j\in\mathcal{J}}\widehat{\lambda}_j\underbrace{\sum_{i\in\mathcal{I}}\hat{\gamma}_{ij}x_{ij}\widehat{\lambda}_j^\frac{-1}{\alpha_i}}_{=1} \\ 
        & ~~~~~ - \sum_{j\in\mathcal{J}}\sum_{i\in\mathcal{I}}\frac{\alpha_i}{1-\alpha_i}\hat{\gamma}_{ij}x_{ij}\widehat{\lambda}_j^\frac{\alpha_i-1}{\alpha_i} \\ 
        & = \sum_{j\in\mathcal{J}}(\lambda_j^*-\widehat{\lambda}_j^*) \\ 
        & ~~~~~ + \sum_{i\in\mathcal{I}}\sum_{j\in\mathcal{J}}\frac{\alpha_i\hat{\gamma}_{ij}x_{ij}}{1-\alpha_i} \left((\lambda_j^*)^\frac{\alpha_i-1}{\alpha_i} - (\widehat{\lambda}_j)^\frac{\alpha_i-1}{\alpha_i} \right).
    \end{split}
\end{equation}

\section{Experimental Details: NVIDIA Sionna}


In our experiments, we utilize the NVIDIA Sionna library~\cite{sionna} to construct standardized 3GPP channel models. 
Specifically, we leverage Sionna's built-in classes \texttt{UMa} and \texttt{UMi} from \texttt{sionna.channel.tr38901}, along with the \texttt{PanelArray} class to define antenna configurations.
We wrap these components into a custom Python function \texttt{get\_channel\_model()}, which selects the appropriate channel model based on the BS transmission power. 
BSs with transmission power between 30 and 36 dBm use the \texttt{UMa} model, while others use \texttt{UMi}.

\begin{python}
import sionna
# A function get channel models
def get_channel_model(num_ofdm_symbols, fft_size, subcarrier_spacing, Fc):
    # Define Resource Grid:
    rg = sionna.ofdm.ResourceGrid(
        num_ofdm_symbols = num_ofdm_symbols,
        fft_size = fft_size,
        subcarrier_spacing = subcarrier_spacing
    )
    # Define BS and UT array:
    bs_array = sionna.channel.tr38901.PanelArray(
        num_rows_per_panel = 1,
        num_cols_per_panel = 1,
        polarization = 'single',
        polarization_type = 'V',
        antenna_pattern = '38.901',
        carrier_frequency = Fc
    )
    ut_array = sionna.channel.tr38901.PanelArray(
        num_rows_per_panel = 1,
        num_cols_per_panel = 1,
        polarization = 'single',
        polarization_type = 'V',
        antenna_pattern = 'omni',
        carrier_frequency = Fc
    )
    channel_model_UMa = sionna.channel.tr38901.UMa(
        carrier_frequency = Fc,
        o2i_model = 'low',
        ut_array = ut_array,
        bs_array = bs_array,
        direction = 'downlink',
        enable_shadow_fading = True,
        enable_pathloss = True,
    )
    channel_model_UMi = sionna.channel.tr38901.UMi(
        carrier_frequency = Fc,
        o2i_model = 'low',
        ut_array = ut_array,
        bs_array = bs_array,
        direction = 'downlink',
        enable_shadow_fading = True,
        enable_pathloss = True,
    )
    return channel_model_UMa, channel_model_UMi
\end{python}

\bibliographystyle{IEEEtran}
\bibliography{main}

\end{document}